\documentclass[a4paper,11pt]{article}
\pdfoutput=1 

\usepackage{jcappub} 

\usepackage[T1]{fontenc} 

\usepackage[utf8]{inputenc}
\usepackage{aas_macros}
\usepackage{amsfonts}
\usepackage{amsmath}
\usepackage{amssymb}
\usepackage{bm}
\usepackage{graphicx}
\usepackage[dvipsnames]{xcolor}
\usepackage{hyperref}
\hypersetup{colorlinks=true,allcolors=teal}

\defcitealias{ps22}{PS22}
\defcitealias{ps23}{PS23}
\defcitealias{cs08}{CS08}

\newcommand{\p}{\ensuremath{\partial}}

\newcommand{\xx}{\ensuremath{\mathbf{x}}}

\renewcommand{\aap}[1]{\ensuremath{\mathbf{a}_{\rm (p)#1}}}

\newcommand{\Mh}{\ensuremath{h^{-1}M_{\odot}}}

\newcommand{\Mpch}{\ensuremath{h^{-1}{\rm Mpc}}}
\newcommand{\hMpc}{\ensuremath{h\,{\rm Mpc}^{-1}}}

\newcommand{\avg}[1]{\ensuremath{\left\langle \,#1\, \right\rangle}}
\newcommand{\e}[1]{\ensuremath{{\rm e}^{#1}}}

\newcommand{\der}{\ensuremath{{\rm d}}}

\newcommand{\dir}{\ensuremath{\delta_{\rm D}}}

\newcommand{\eqn}[1]{equation~\eqref{#1}}
\newcommand{\eqns}[1]{equations~\eqref{#1}}
\newcommand{\ph}[1]{\phantom{#1}}

\newcommand{\beq}{\begin{equation}}
\newcommand{\eeq}{\end{equation}}
\newcommand{\Cal}[1]{\ensuremath{\mathcal{#1}}}

\newcommand{\xiell}[1]{\ensuremath{\xi_{\rm NL}^{(#1)}}}
\newcommand{\xiellprop}[1]{\ensuremath{\xi_{\rm prop}^{(#1)}}}
\newcommand{\xiellmc}[1]{\ensuremath{\xi_{\rm MC}^{(#1)}}}

\newcommand{\xiellsim}[1]{\ensuremath{\hat \xi}^{(#1)}}
\newcommand{\Dellsq}[1]{\ensuremath{\Delta^{(#1)2}_{\rm NL}}}
\newcommand{\Dellpropsq}[1]{\ensuremath{\Delta^{(#1)2}_{\rm prop}}}
\newcommand{\Dellmcsq}[1]{\ensuremath{\Delta^{(#1)2}_{\rm MC}}}
\newcommand{\DellLsq}[1]{\ensuremath{\Delta^{(#1)2}_{\rm L}}}
\newcommand{\sigv}{\ensuremath{\sigma_{\rm v}}}
\newcommand{\Pell}[1]{\ensuremath{\mathcal{P}_{#1}}}

\newcommand{\qq}{\ensuremath{\mathbf{q}}}
\newcommand{\vv}{\ensuremath{\mathbf{v}}}
\newcommand{\zz}{\ensuremath{\mathbf{z}}}
\renewcommand{\ss}{\ensuremath{\mathbf{s}}}

\newcommand{\kk}{\ensuremath{\mathbf{k}}}

\title{\boldmath Scale-dependent bias and mode coupling in redshift-space clustering near the BAO scale}

\author[a]{Aseem Paranjape}
\author[b,c]{and Ravi K. Sheth}

\affiliation[a]{Inter-University Centre for Astronomy \& Astrophysics,\\ Ganeshkhind, Post Bag 4, Pune 411007, India}
\affiliation[b]{Center for Particle Cosmology, University of Pennsylvania,\\ 209 S. 33rd St., Philadelphia, PA 19104, USA}

\affiliation[c]{The Abdus Salam International Center for Theoretical Physics,\\ Strada Costiera, 11, Trieste 34151, Italy}
 
\emailAdd{aseem@iucaa.in}
\emailAdd{shethrk@physics.upenn.edu}

\abstract{
The baryon acoustic oscillation (BAO) feature in the 2-point clustering of biased tracers in redshift space can be described in a model-agnostic manner, relying only on the assumption that nonlinear growth approximately smears this feature with a Gaussian kernel sourced by gravitationally driven bulk flows as in the Zel'dovich approximation. 
An explicit model that demonstrated this in recent work did not account for two physical effects that are very likely observationally relevant in the context of ongoing surveys, namely, the scale-dependence of linear Lagrangian density and velocity bias and the effects of mode coupling. 
We rectify this shortcoming in this paper by showing that a simple model including these effects is able to accurately describe the multipoles of the 2pcf of realistic tracer samples at BAO scales. 
Our results indicate that the effects of scale-dependent bias will be important to model for surveys such as DESI, while those of mode coupling are relatively less significant. 
Our model for scale-dependent bias and mode coupling, which is motivated by model-agnostic arguments from peaks theory and the Zel'dovich approximation, lies in the class of `Laplace-Gauss' expansions, making it straightforward to incorporate these effects in the model-agnostic inference framework mentioned above. 
}


\begin{document}
\maketitle
\flushbottom

\section{Introduction}
\label{sec:intro}
The baryon acoustic oscillation (BAO) feature in the clustering of biased tracers such as galaxies is one of the primary science drivers in current, upcoming and planned surveys of large-scale structure \cite{eisenstein+05,cole+05,anderson+12,anderson+14,alam+17}. This is not only from the viewpoint of increasingly precise constraints on parameters in the Lambda-cold dark matter ($\Lambda$CDM) model, but also in the hunt for signatures of departures from the standard model \cite{bottaro+24,mhc24,pmp24}. Traditional methods \cite{cuesta+16,beutler+17,blomqvist+19,dumasdesbourbouz+20,gil-marin+20,extractor2020,abbott+22} exploit the information content of the BAO feature in 2-point clustering (in configuration or Fourier space) by assuming templates based on a fiducial $\Lambda$CDM model and constraining parameters describing variations around this model. Field-level inference techniques are now being explored to maximise information recovery from the BAO feature within $\Lambda$CDM \cite{bst24,bst25}. Recent work has also shown how this information might be repackaged in model-agnostic analyses that do not rely on $\Lambda$CDM templates \cite{LP2016,LPmocks,LPboss,nsz21a,nsz21b,ps22}. 

In a recent work \cite[][hereafter, PS23]{ps23}, we showed how the shape of the BAO feature in redshift space, as characterised by the multipoles of the 2-point correlation function (2pcf) of galaxies, can be described by a model that \emph{does not} assume the $\Lambda$CDM framework. Rather, we assume that the Zel'dovich approximation is a good description of the nonlinear motion of matter at large scales. In this case, the BAO feature is smeared by an approximately Gaussian kernel \cite{bharadwaj96,cs06b} whose width \sigv\ is a parameter of cosmological interest. \citetalias{ps23} showed that this model (which we recapitulate below) can successfully describe the clustering of galaxies in existing surveys and provide meaningful constraints on \sigv\ as well as the linear growth rate $f$, \emph{without} assuming a $\Lambda$CDM cosmology. However, the model as stated in \citetalias{ps23} was unable to accurately recover these parameters when making projections for ongoing surveys such as the one by the Dark Energy Spectroscopic Instrument (DESI) collaboration \cite{DESI}. This is particularly relevant in the context of the recent early results from DESI hinting at possible departures from the standard model \cite{DESI2024-cosmo}.

Our aim in the present work is to improve upon the \citetalias{ps23} model by rectifying two of its main shortcomings, namely, the lack of scale-dependence of the bias of the underlying tracers and the lack of mode coupling effects expected due to nonlinear growth. On BAO scales, the former is expected to be more important than the latter \cite{dcss10,bkBAO}.  Nevertheless, we will show that both of these effects can be incorporated into an extension of the \citetalias{ps23} model adding only minimal assumptions: 
(a) that the distribution and motion of point tracers such as galaxies is modified compared to the matter distribution along the lines suggested by peaks theory, and 
(b) that, on BAO scales, the dominant contribution to the mode-coupling effect comes from a dipole-like contribution whose form is constrained by Galilean invariance \cite{mpp24}. 
Our focus in this paper is to demonstrate that our resulting model of redshift space multipoles of the 2pcf including scale-dependent bias and mode coupling provides an accurate description of measurements of these quantities in $\Lambda$CDM simulations for realistic tracer populations. We leave the application of our setup in a full model-agnostic inference exercise to a forthcoming publication.

The paper is structured as follows. In section~\ref{sec:model} we recapitulate the model from \citetalias{ps23} and then describe our modifications to include the effects of scale-dependent bias and mode coupling. We describe the simulations we use, along with our analysis for validating the model, in section~\ref{sec:sims-analysis}. Section~\ref{sec:results} presents our results for the base model we propose in this work, and also reports the results of variations around this model as a means of testing its robustness. We conclude in section~\ref{sec:conclude}.
Although built using Fourier space ingredients, our model is intended to be accurate at large scales in configuration space. As a sanity check, however, in Appendix~\ref{app:FourierMC} we also study its accuracy in Fourier space at low values of wavenumber $k$.

\section{Model}
\label{sec:model}
Our starting point is a description of the non-linear clustering of biased tracers near the BAO feature in redshift space that is motivated by the Zel'dovich approximation written in the language of renormalized perturbation theory \cite{cs06a,cs06b,cs08}. Here, one writes the non-linearly evolved galaxy 2pcf in redshift space, $\xi_{\rm NL}(\ss)$, as a sum of two contributions,
\beq
\xi_{\rm NL}(\ss) = \xi_{\rm prop}(\ss) + \xi_{\rm MC}(\ss)\,,
\label{eq:xiNL-generic}
\eeq
where the `propagator' term $\xi_{\rm prop}(\ss)$ accounts for the effect of bulk flows on the large-scale tracer correlations, while the mode coupling term $\xi_{\rm MC}(\ss)$ additionally accounts for non-linearities at smaller scales.
\citetalias{ps23} discussed a simple model for $\xi_{\rm prop}(\ss)$ (which we recapitulate below) and ignored $\xi_{\rm MC}(\ss)$. Below, we extend their model to account for the effects of, both, scale-dependent bias and mode coupling. In particular, we argue that the mode coupling term $\xi_{\rm MC}(\ss)$ is expected to be  related to the propagator term $\xi_{\rm prop}(\ss)$.


\subsection{Smearing due to displacements}
If $\xi_{\rm L}(\ss)$ is the linearly evolved galaxy 2pcf in redshift space, 
and we focus on large separations $s$ close to the BAO feature, then the effect of bulk flows in the propagator term of \eqn{eq:xiNL-generic} is to smear the linear 2pcf according to \citep{bharadwaj96,cs06b,cs08}
\beq
\xi_{\rm prop}(\ss) \approx \int\der^3s^\prime\,\xi_{\rm L}(\ss^\prime)\,\Cal{N}\left(\ss-\ss^\prime;\,\mathbf{\Sigma}\right)\,,
\label{eq:smearing-redshiftspace}
\eeq
where $\Cal{N}(\xx;\mathbf{\Sigma})$ denotes a 3-dimensional Gaussian distribution in \xx\ having zero mean and covariance matrix $\mathbf{\Sigma}$. 
This matrix is anisotropic in redshift space, with variances $\sigma_\parallel^2$ and $\sigma_\perp^2$ along and perpendicular to the line-of-sight, respectively, given by \citep{ZeldovichRSD,ds10,ppvv15}
\begin{align}
\sigma_\parallel^2 = 2\sigv^2\,(1+f)^2\,\quad {\rm and}\quad 
\sigma_\perp^2 = 2\sigv^2\,,
\label{eq:sigpar-sigperp}
\end{align}
where $f\equiv\der\ln D/\der\ln a$ is the usual linear growth rate and we defined the linear theory 1-dimensional, single-particle velocity dispersion \sigv\
\beq
\sigv^2\equiv \frac13\int\der\ln k\,k^{-2}\,\Delta^2_{\rm lin}(k)\,,
\label{eq:sigv-def}
\eeq
with $\Delta^2_{\rm lin}(k)=k^3P_{\rm lin}(k)/(2\pi^2)$ being the dimensionless linear theory matter power spectrum (whose redshift dependence is suppressed). The large-scale redshift-space 2pcf 
$\xi_{\rm L}(\ss)$ appearing in \eqn{eq:smearing-redshiftspace} is
\beq
\xi_{\rm L}(\ss) = \int\frac{\der^3k}{(2\pi)^3}\,\e{i\kk\cdot\ss}\,b^2\left(1+\beta\mu_k^2\right)^2\,P_{\rm lin}(k)\,,
\label{eq:xiL-redshift}
\eeq
where $b$ is the large-scale Eulerian bias of the sample and $\beta\equiv f/b$ \citep{kaiser87}.

This gives the expression in equation~(17) of \citetalias{ps23} for the non-linear power spectrum multipoles\footnote{\citetalias{ps23} referred to this as their `exact' model for the multipoles of the \emph{full} power spectrum \Dellsq{\ell}.} \Dellpropsq{\ell}:
\begin{align}
\Dellpropsq{\ell}(k) 
&= \Delta_{\rm lin}^2(k)\,b^2\,\e{-k^2\sigv^2}\,(2\ell+1)
\int_{-1}^1\frac{\der\mu_k}{2}\,\Pell{\ell}(\mu_k)\left(1+\beta\mu_k^2\right)^2\e{-K^2\mu_k^2}\,,
\label{eq:Delta(ell)2-def}
\end{align}
where
\beq
K^2 \equiv k^2\sigv^2\,f(f+2)\,,
\label{eq:K2-def}
\eeq
and $\Cal{P}_\ell(\mu)$ is the Legendre polynomial of degree $\ell$. Our goal here is to extend this model to include the effects of scale-dependent density and velocity bias along with mode coupling.

\subsection{Modelling scale-dependent bias}
\label{subsec:scaledepbias}
To include the effects of scale-dependence of the density and velocity bias at linear order, we make the replacement
\begin{align}
b+f\mu_k^2 &\to b_{\rm Lag}(k) + b_{\rm vel}(k)(1 + f\mu_k^2) \equiv b\,B(k,\mu_k)\,.
\label{eq:scaledepbias}
\end{align}
We discuss our model for $b_{\rm Lag}(k)$ and $b_{\rm vel}(k)$ below. Our starting point is the result that, at fixed halo mass $m$, the Lagrangian linear density and velocity bias factors are very well approximated by the peaks theory form \cite{ds10,bds15,bkBAO} 
\begin{align}
b_{\rm Lag}(k|m) &= \left[b_{10}(m) + b_{01}(m)\,k^2R_{\rm pk}^2\right]\e{-k^2R_{\rm h}^2/2}\,,
\label{eq:bLag(k|m)} \\
b_{\rm vel}(k|m) &= \left[1 - k^2R_{\rm pk}^2\right]\e{-k^2R_{\rm h}^2/2}\,, 
\label{eq:bvel(k|m)} 
\end{align}
where $R_{\rm h}=(3m/4\pi\bar\rho)^{1/3}/\sqrt{5}$ is the Lagrangian scale\footnote{The factor of $\sqrt{5}$ accounts for the fact that the Lagrangian scale is generally associated with a tophat filter in real space, while we are working with Gaussian filters; see the discussion in \cite{phs18}.} corresponding to mass $m$ and $R_{\rm pk}$ is the characteristic peaks scale for mass $m$, given by
\beq
R_{\rm pk}(m) \equiv \frac{\sigma_{0}(R_{\rm h})}{\sigma_1(R_{\rm h})}\,,
\eeq
where the $z=0$ spectral integrals $\sigma^2_j(R)$ are given by
\beq
\sigma_j^2(R) \equiv \int\der\ln k\,\Delta^2_{\rm lin}(k)\,k^{2j}\,\e{-k^2R^2}\,.
\label{eq:spectral}
\eeq
In \eqn{eq:bLag(k|m)}, $b_{10}(m)$ is the peak-background split contribution to linear Lagrangian density bias and $b_{01}(m)$ is related to $b_{10}(m)$ through a consistency relation of the form
\beq
b_{10}(m) + b_{01}(m) = \frac{\delta_{\rm crit}(m)}{\sigma_0(R_{\rm h})^2}\,,
\eeq
where $\delta_{\rm crit}(m)$ is the linearly extrapolated critical density for collapse, whose canonical value at $z=0$ in the spherical collapse model is $\simeq1.686$, but which can have an effective mass dependence due to effects of non-sphericity \cite{smt01,st02} or tides \cite{cphs17}. 

The detailed mass dependence of the factors $b_{10}$ and $b_{01}$ is quite model dependent. Moreover, galaxy samples typically do not correspond to narrow -- or even well-defined -- ranges of halo mass. In principle, we then need a model that is flexible enough to capture not only the inherent uncertainties in modelling the fixed-mass coefficients above, but also the effects of averaging over halo mass. The structure in \eqns{eq:bLag(k|m)}-\eqref{eq:bvel(k|m)} motivates a model for $b_{\rm Lag}(k)$ and $b_{\rm vel}(k)$ given by
\begin{align}
b_{\rm Lag}(k) &= \left[b-1 + b_{01}k^2R_{\rm p}^2\right] \e{-k^2R_\ast^2/2}\,,
\label{eq:bLag-model} \\
b_{\rm vel}(k) &= \left[1 -B_{v} k^2R_{\rm p}^2\right]  \e{-k^2R_\ast^2/2}\,,
\label{eq:bvel-model}
\end{align}
where $b_{01}$, $B_{v}$ and $R_\ast$ are free parameters and $R_{\rm p}$ is a fixed pivot scale.
This in turn leads to the following expression for $B(k,\mu_k)$ (equation~\ref{eq:scaledepbias})
\begin{align}
B(k,\mu_k) &= b^{-1}\left[b-1 + b_{01}k^2R_{\rm p}^2 + \left(1 - B_{v}k^2R_{\rm p}^2\right)(1 + f\mu_k^2)\right]  \e{-k^2R_\ast^2/2} \notag\\
&= b^{-1}\left[b + (b_{01}-B_{v})R_{\rm p}^2k^2 + f\mu_k^2(1-B_{v}R_{\rm p}^2k^2)\right] \e{-k^2R_\ast^2/2} \notag\\
&\equiv \left[1 + B_1k^2 R_{\rm p}^2 + \beta\mu_k^2\left(1 - B_vk^2 R_{\rm p}^2\right)\right] \e{-k^2R_\ast^2/2}\,,
\label{eq:B(k,mu)}
\end{align}
where we defined $B_1\equiv(b_{01}-B_v)/b$. Our model for scale-dependent bias thus has $3$ free parameters in addition to the large-scale bias $b$: $R_\ast$ represents a characteristic Lagrangian scale of the tracer population, while $B_1$ and $B_v$ capture the additional peaks-theory-like effects of scale-dependence. The pivot scale $R_{\rm p}$ only serves to make $B_1$ and $B_v$ dimensionless; throughout this work we fix its value to $R_{\rm p}=2.5\Mpch$.
The primary advantage of this model is that it falls within the class of `Laplace-Gauss' expansions discussed by \citetalias{ps23}, which allow for straightforward implementations of model-agnostic inference as discussed by those authors. We return to this point in section~\ref{sec:conclude}.
We will test this model below using redshift-space clustering measurements for mass-thresholded halo samples in $N$-body simulations of $\Lambda$CDM. 

\subsection{Modelling mode coupling}
To model the potentially anisotropic effects of mode coupling (MC), we follow \cite[][CS08]{cs08} who argue that the leading order MC effect near the BAO feature in real space is of the form $(\sim\der\xi_{\rm lin}/\der\ln r)\,\bar\xi_{\rm lin}$. Further noting that the volume averaged 2pcf $\bar\xi_{\rm lin}$ is featureless and relatively flat across the BAO feature, \citetalias{cs08} motivate modelling the monopole of the redshift space 2pcf of biased tracers by adding a term proportional to $\der\xi_{\rm lin}(s)/\der\ln s$ to the 2pcf of our previous Zel'dovich smearing approximation (their equation 34). 

We discuss how to extend this to biased, redshift space distorted tracers, and how strongly this depends on $\Lambda$CDM, in Appendix~\ref{app:mc}. Importantly, this discussion shows that, similarly to the calculation by \citetalias{cs08} for dark matter, the dominant contribution to the mode coupling term for tracers with scale-dependent bias is \emph{also} expected to be approximately proportional to a derivative of the propagator term. Motivated by this, we consider the following ansatz for $\xi_{\rm NL}(\ss)$ that includes anisotropic effects of, both, mode coupling as well as scale-dependent bias:
\begin{align}    
\xi_{\rm NL}(\ss) &=
\int\der^3s^\prime\,\xi_{\rm L}(\ss^\prime|0)\,\Cal{N}\left(\ss-\ss^\prime;\,\mathbf{\Sigma}\right) + A_{\rm MC}\,\frac{\p\xi_{\rm L}(\ss|R_{\rm MC})}{\p\ln s}\,,
\label{eq:xiNL-sdbmc}
\end{align}
with
\beq
\xi_{\rm L}(\ss|R) \equiv \int\frac{\der^3k}{(2\pi)^3}\,\e{i\kk\cdot\ss}\,b^2\,B(k,\mu_k)^2\,\e{-k^2R^2}\,P_{\rm lin}(k)\,.
\label{eq:xiLbk-redshift}
\eeq
%
Note that the expression for $\xi_{\rm L}(\ss|R)$ is different from the one for $\xi_{\rm L}(\ss)$ in \eqn{eq:xiL-redshift} not only due to the Gaussian smearing with scale $R$ but also the scale-dependent bias factor $B(k,\mu_k)$. The first integral in \eqn{eq:xiNL-sdbmc} is our model for $\xi_{\rm prop}(\ss)$, while the second term is our approximation for $\xi_{\rm MC}(\ss)$. Appendix~\ref{app:mc} would actually suggest setting $\xi_{\rm MC}(\ss) = A_{\rm MC}\,\der\xi_{\rm prop}(\ss)/\der\ln s$, so that the smearing appearing in this term is anisotropic, being derived from that in $\xi_{\rm prop}(\ss)$. There are, however, several approximations made before arriving at this proportionality in Appendix~\ref{app:mc} and, strictly,  the pre-factor $A_{\rm MC}$ should also not be a constant. To approximately account for these differences, we have chosen to simplify the ansatz and only use an isotropic smearing when describing $\xi_{\rm MC}(\ss)$, while allowing the smearing scale $R_{\rm MC}$ to be a free parameter and not necessarily equal to \sigv\ (which appears in the anisotropic kernel in $\xi_{\rm prop}(\ss)$). We will see below, however, that our comparisons with numerical simulations strongly support setting $R_{\rm MC}\simeq\sigv$ in \eqn{eq:xiNL-sdbmc}, which builds confidence in our approximations.


\vskip 0.1in
\noindent
Thus, our model of nonlinear redshift-space clustering including scale-dependent bias and mode coupling, which we refer to as \emph{sdbmc} below, contains 5 free parameters $\{B_1,B_v,A_{\rm MC},R_{\rm MC},R_\ast\}$ apart from the large-scale bias $b$, the velocity dispersion \sigv\ and the growth rate $f$.
Here, $\{B_1,B_v,R_\ast\}$ describe the effects of scale-dependent bias and $\{A_{\rm MC},R_{\rm MC}\}$ describe mode coupling. 
When these 5 parameters are set to zero, the model reduces to the one discussed by \citetalias{ps23}, which we refer to as \emph{no sdbmc} below.

Throughout this work, we will set \sigv\ and $f$ to the values inferred from the appropriate cosmological model, mentioned in section~\ref{subsec:sims}. In a full-fledged inference exercise, of course, \sigv\ and $f$ would be left free; we will study this in a forthcoming publication. 
For simplicity, below we will fix $b$ and $R_\ast$ to the values inferred from the large-scale clustering and mass distribution of the halo sample and only vary $\{B_1,B_v,A_{\rm MC},R_{\rm MC}\}$. When considering galaxy samples, which need not correspond to well-defined halo mass cuts, the values of $b$ and $R_{\ast}$ might also need to be varied,  possibly with well-motivated priors (see, e.g., the discussion in \citetalias{ps23} and also below). 

\subsection{Configuration space multipoles}
We will compare this model with 2pcf measurements in configuration space, near the BAO scale, and consequently need the multipole moments of the preceding expressions.
A calculation similar to that leading from equation~(10) to equation~(16) in \citetalias{ps23} gives
\beq
\xiell{\ell}(s) = \xiellprop{\ell}(s) + \xiellmc{\ell}(s)\,,
\label{eq:xiell}
\eeq
where
\begin{align}
\xiellprop{\ell}(s) &= i^\ell\int\der\ln k\,\Dellpropsq{\ell}(k)\,j_\ell(ks)\,, 
\label{eq:xiellprop}\\
\xiellmc{\ell}(s) &= A_{\rm MC}\,i^\ell\int\der\ln k\,\DellLsq{\ell}(k)\left[\ell\,j_\ell(ks) - ks\,j_{\ell+1}(ks)\right]\,, \label{eq:xiellmc}
\end{align}
with
\begin{align}
\Dellpropsq{\ell}(k) &\equiv \Delta_{\rm lin}^2(k)\,b^2\,\e{-k^2\sigv^2}\,(2\ell+1)
\int_{-1}^1\frac{\der\mu_k}{2}\,\Pell{\ell}(\mu_k)\,B(k,\mu_k)^2\e{-K^2\mu_k^2}\,, \label{eq:Dellprop} \\
\DellLsq{\ell}(k) &\equiv \Delta_{\rm lin}^2(k)\,b^2\,\e{-k^2R_{\rm MC}^2}\,(2\ell+1)
\int_{-1}^1\frac{\der\mu_k}{2}\,\Pell{\ell}(\mu_k)\,B(k,\mu_k)^2\,, \label{eq:DellL}
\end{align}
and we used the identity 
\beq
x\,\der j_\ell(x)/\der x = \ell\,j_\ell(x) - x\,j_{\ell+1}(x) \,,
\label{eq:sphbessell-derivrecur}
\eeq
in writing \eqn{eq:xiellmc}. 
In the limit of scale independent bias ($B(k,\mu_k)\to 1+\beta\mu_k^2$), 
$\Dellpropsq{\ell}(k)$ reduces to \eqn{eq:Delta(ell)2-def} (i.e., equation~(17) of \citetalias{ps23}), while $\DellLsq{\ell}(k)$ reduces to the Kaiser multipoles, $\DellLsq{\ell}(k)\to\Delta_{\rm lin}^2(k)b^2\chi_\ell(\beta)$. 
This model can also be compared with multipole measurements in Fourier space; we provide the necessary conversions in Appendix~\ref{app:FourierMC}. In fact, to compute $\xiellmc{\ell}(s)$ in practice, we first evaluate \eqns{eq:D0MC}-\eqref{eq:D4MC} in Fourier space and then perform an integral over $k$.

\section{Simulations and Analysis}
\label{sec:sims-analysis}
\subsection{Simulations}
\label{subsec:sims}
For the results in the main text, we rely on 20 realisations of the HADES $N$-body simulations \cite{hades}. Each run evolved $512^3$ particles in a $(1h^{-1}{\rm Gpc})^3$ volume using a flat $\Lambda$CDM cosmology with parameters $\Omega_{\rm m}=0.3175$, $\Omega_{\rm b}=0.049$, $h=0.6711$ $n_{\rm s}=0.9624$, $\sigma_8=0.833$, with a particle mass of $m_{\rm part}=6.56\times10^{11}\Mh$. Specifically, we use mass-weighted\footnote{Although the mass-weighting was implemented by the authors of \cite{nikakhtar+23} to aid their reconstruction analysis, in our case it has the desirable effect of approximating the impact of satellite galaxies in the sample.} measurements of the configuration space multipoles of the 2pcf $\xiellsim{\ell}$ for $\ell=0,2,4$ presented by \cite{nikakhtar+23} using halo catalogs at $z=0$, with halos identified using the Friends-of-Friends algorithm. The halo sample in each realisation was defined using a mass threshold of $m\geq20\,m_{\rm part}\simeq1.3\times10^{13}\Mh$. The combined volume of the 20 HADES realisations is similar to the effective volume of the LRG sample in the DESI survey \cite{DESI}. The large-scale clustering strength of this sample (see below) is $b\simeq1.95$, also comparable to that expected for DESI LRGs. The effective number density of the sample\footnote{Due to mass-weighting, the number density is not simply the ratio of the number of tracers to the box volume $V$. Rather, it must be calculated using $\bar n^{-1} =  \sum_im_i^2/V / \left(\sum_im_i/V\right)^2$, where $m_i$ is the mass of the $i^{\rm th}$ halo and the sum is over all halos in the sample. This leads to $\bar n^{-1}\simeq \, 9000 (\Mpch)^3$ \cite[e.g., the dashed line in fig.~5 of][]{nikakhtar+23}.} is $\bar n\simeq1.1\times10^{-4}\,(\hMpc)^3$, about a factor 6 smaller than the expected DESI LRG density. Note, however, that the HADES measurements are at $z=0$ while DESI LRGs are expected to lie at $z\sim0.7$, so a direct comparison cannot be made. The growth factor for this cosmology and redshift is $f=0.53$, while the linear velocity dispersion is $\sigv=6.0\Mpch$. 

For the results in Appendix~\ref{app:FourierMC}, we rely on the MINERVA $N$-body simulations described by Grieb \emph{et al.} \cite{grieb2016}. These evolved $1000^3$ particles in a $(1.5h^{-1}{\rm Gpc})^3$ volume using a flat $\Lambda$CDM cosmology with parameters $\Omega_{\rm m}=0.285$, $\Omega_{\rm b}=0.04606$, $h=0.695$ $n_{\rm s}=0.9632$, $\sigma_8=0.828$. Grieb \emph{et al.} further constructed mock galaxy catalogs using a halo occupation distribution (HOD) model in halos identified at $z=0.57$, designed to mimic the CMASS sample of the BOSS survey \cite{BOSSDR12-FinalData} and having $b=2.005$ and $\bar n = 4\times10^{-4}\,(\hMpc)^3$. We use the Fourier space measurements of power spectrum multipoles presented in fig.~3 of \cite{grieb2016}. The corresponding configuration space measurements are not available to us. The growth factor and linear velocity dispersion for this cosmology and redshift are $f=0.76$ and $\sigv=4.6\Mpch$.

\subsection{Gauss-Poisson covariance matrix}
\label{subsec:GPcov}
As in \citetalias{ps23}, we work in the Gauss-Poisson approximation, where the covariance matrix
\beq
C^{\ell\ell^\prime}_{ij} \equiv \avg{\xiell{\ell}(s_i)\xiell{\ell^\prime}(s_j)} - \avg{\xiell{\ell}(s_i)}\avg{\xiell{\ell^\prime}(s_j)}
\label{eq:covmat-def}
\eeq
can be written as \citep[e.g.,][]{grieb2016},
\begin{equation}
C^{\ell\ell^\prime}_{ij} = \frac{i^{\ell_1 + \ell_2}}{2\pi^2}\int dk\,k^2\,{\bar j}_{\ell_1}(ks_i)\,{\bar j}_{\ell_2}(ks_j) \,\sigma^2_{\ell_1\ell_2}(k)\,.
\label{eq:C_ell1ell2(si,sj)}
\end{equation}
Here
\begin{align}
\sigma^2_{\ell_1\ell_2}(k) &= \frac{(2\ell_1 + 1)(2\ell_2 + 1)}{V_{\rm sur}/2}\notag\\
&\ph{V_{sur}/2}
\times\int_{-1}^{1} \frac{d\mu}{2}\,
\left[P(k,\mu) + \frac{1}{\bar{n}}\right]^2\,\Pell{\ell_1}(\mu)\,\Pell{\ell_2}(\mu) 
\label{eq:sig2_ell1ell2(k)}
\end{align}
for a survey of volume $V_{\rm sur}$ with observed tracer number density $\bar n$ and 
\beq
{\bar j}_{\ell}(ks_i) = \frac{4\pi}{V_i}\int ds\,s^2\,j_\ell(ks)\,W_i(s)
\label{eq:jellbar}
\eeq
with
\beq
V_i = 4\pi \int ds\,s^2\,W_i(s)\,,
\eeq
where $W_i(s)$ describes the shape of a window over which $j_\ell$ has been averaged.  E.g., for the tophat bins of width $\Delta s$ centered on $s_i$ that we use, $W_i(s) = 1$ if $s_i-\Delta s/2 \le s\le s_i + \Delta s/2$ and zero otherwise, and the integral which defines $\bar{j}_\ell$ can be done analytically. 

For the configuration-space results in the main text, we further scale the Gauss-Poisson covariance matrix so as to maintain its correlation structure while agreeing with the diagonal errors on $\xiellsim{\ell}(s)$ from the simulation measurements.
For the Fourier space results in Appendix~\ref{app:FourierMC}, we integrate $\sigma^2_{\ell_1\ell_2}(k)$ over appropriate bins in $k$, rather than the Bessel-weighted integrals in \eqn{eq:sig2_ell1ell2(k)}, and use the Gauss-Poisson matrix directly. Neither of these choices significantly impacts our final results or conclusions.

We perform parameter inference using the Monte Carlo Markov Chain (MCMC) technique. Below, we use the publicly available \textsc{Cobaya} \citep{tl19-cobaya,tl21-cobaya}\footnote{\url{https://cobaya.readthedocs.io/}} and \textsc{GetDist} \citep{lewis19},\footnote{\url{https://getdist.readthedocs.io/}} packages to implement the MCMC and display results, respectively, discarding the first $30\%$ of the samples as burn-in. We assume a Gaussian likelihood throughout. Wherever needed, we generate cosmological transfer functions for matter fluctuations using the \textsc{class} code \cite{class-I,class-II}.\footnote{\url{http://class-code.net/}}

\section{Results}
\label{sec:results}

\subsection{Base model}
\label{subsec:basemodel}
We now attempt to describe the configuration space multipole moment measurements $\xiellsim{\ell}(s)$ at $z=0$ at BAO scales for $\ell=0,2,4$ in the HADES simulations presented by \cite{nikakhtar+23}, using the \emph{sdbmc} model. In Appendix~\ref{app:FourierMC}, we report the results of a similar exercise to describe the Fourier space multipole measurements reported by \cite{grieb2016} in the MINERVA simulations at $z=0.57$.

As mentioned above, for this exercise we set $R_\ast=2.5\Mpch$ to approximately match the characteristic mass of the sample discussed in section~\ref{subsec:sims}. We have checked that small variations in this value do not affect the details or quality of our fits below. 
In general, the scale as well as shape of the filter could be varied, e.g., by expanding around a Gaussian shape in powers of $k^2$.\footnote{Recall that the pivot scale is also fixed to $R_{\rm p}=2.5\Mpch$; unlike $R_\ast$, however, this scale is never varied as a free parameter.}
To proceed with the analysis, we must deal with the fact that the Gauss-Poisson covariance matrix depends on the model for the power spectrum of the biased tracers.
In the first pass, we treat the standard error on the mean of $\xiellsim{\ell}(s)$ across 20 HADES realizations as uncorrelated measurement errors so as to get an estimate of the values of $B_1$, $B_v$, $A_{\rm MC}$ and $R_{\rm MC}$ that best describe the measurements. Of these, the values of $B_1$ and $B_v$ are then used in a Gauss-Poisson estimate of the measurement covariance for the simulation box, while keeping $A_{\rm MC}=0=R_{\rm MC}$ for the covariance calculation, to improve the estimate of these parameters using another MCMC run. The resulting best fitting parameter values are then used in a second estimate of the Gauss-Poisson, which is used in a final parameter inference exercise. We find converged results at this point for, both, the best fit values and errors as well as the goodness of fit. The differences between the initial and final estimates of the best fit are also small and statistically insignificant. 
Below, we report only the results using the full, converged covariance matrix.
Our choice of not including MC terms in the covariance estimate is justified \emph{post hoc} by the fact that the MC contribution below is estimated to be small.

\begin{figure*}[h]
\centering
\includegraphics[width=0.49\textwidth]{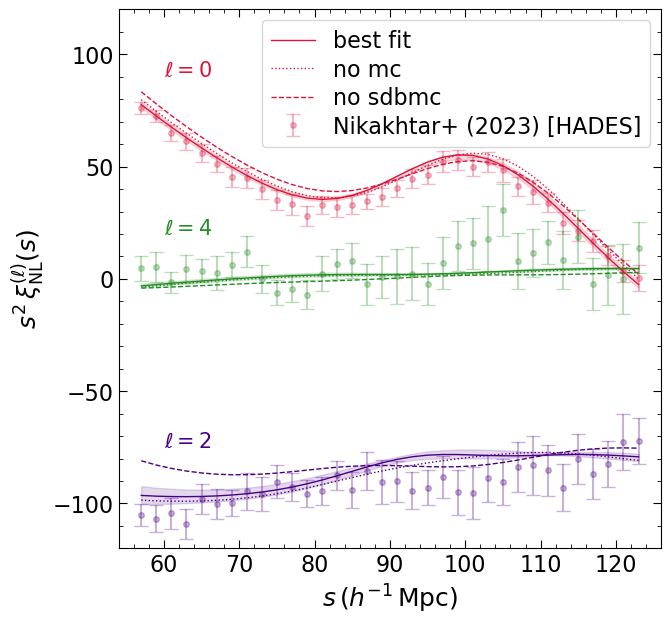}
\includegraphics[width=0.5\textwidth]{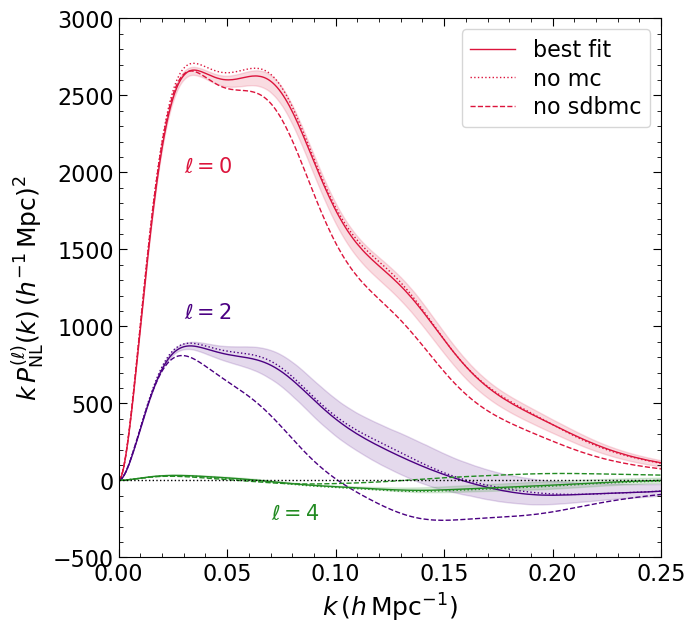}
\caption{
\emph{(Left panel):} Comparison of the HADES configuration-space measurements from \cite{nikakhtar+23} (points with error bars) with the best fitting \emph{sdbmc} model (equation~\ref{eq:xiell}; solid curves with error bands). For comparison, the dashed curves show the \emph{no sdbmc} model setting the scale-dependent and mode coupling terms to zero, while the dotted curves show the result when setting only the mode coupling terms to zero ($A_{\rm MC}=0$) with other parameters set to their best-fit values. \emph{(Right panel):} Fourier-space results corresponding to the configuration-space ones from the left panel, formatted identically.
This analysis used $b=1.95$ and $R_\ast=2.5\Mpch$, with cosmological parameters set to the ones used in the HADES simulations (see text for details).}
\label{fig:sdbmc-stats}
\end{figure*}

In principle, since we know the cosmology for the HADES simulation and because the tracer sample is defined by a simple threshold on halo mass, we can estimate the value of $b$ (the large-scale Eulerian bias of the sample) using a mass-weighted calculation with the fitting functions for linear bias and halo mass function from \cite{Tinker10} and \cite{Tinker08}, respectively, which leads to $b=2.08$. These fitting functions are expected to have a systematic error of about $\sim5\%$, however. We therefore performed an MCMC analysis varying $b$ along with the 4 \emph{sdbmc} parameters, finding that $b=1.947^{+0.042}_{-0.031}$ (best fit with central $68\%$ confidence interval) with only weak degeneracies with the other parameters. This is statistically consistent with the expectation from the fitting functions from the literature, when accounting for the $\sim5\%$ error mentioned above, and is also a more precise estimate ($\lesssim2\%$ error). In the following, therefore, we fix the large-scale bias to $b=1.95$.

\begin{figure}[h]
\centering
\includegraphics[width=0.7\textwidth]{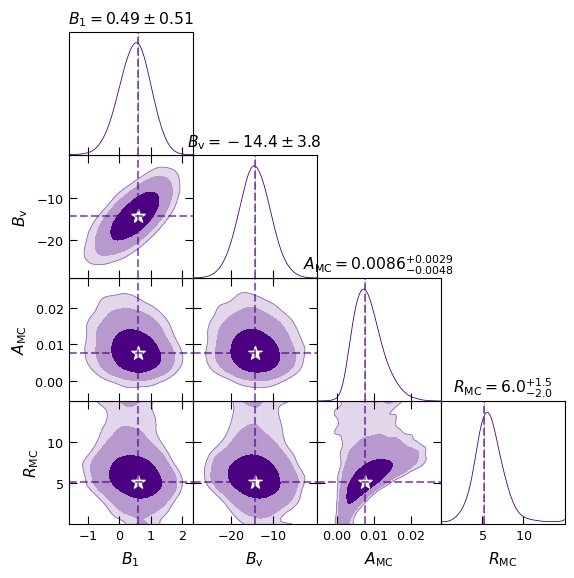}
\caption{Constraints on \emph{sdbmc} model parameters using the HADES simulation measurements. The contours show the $68\%$, $95\%$ and $99\%$ confidence regions, and the dashed lines intersecting at the white stars show the best-fit parameter combination. This analysis used $b=1.95$ and $R_\ast=2.5\Mpch$, with cosmological parameters set to the ones used in the HADES simulations (see text for details).}
\label{fig:sdbmc-contours}
\end{figure}

Fig.~\ref{fig:sdbmc-stats} shows the best fit model describing the HADES simulation measurements, with Fig.~\ref{fig:sdbmc-contours} showing the corresponding contour plots of the parameter constraints. The best fitting parameter combination and $\chi^2$ are reported in Table~\ref{tab:sdbmc-bestfit}. The model is clearly a very good fit. 
We have also checked
that the data and the best fitting \emph{sdbmc} model exclude the \emph{no sdbmc} model with substantial statistical significance: the $\chi^2$ value for the \emph{no sdbmc} model is larger than that for the best-fitting \emph{sdbmc} by $\Delta\chi^2\sim33$, and the confidence contours in Fig.~\ref{fig:sdbmc-contours} exclude $A_{\rm MC}=0$ at $>95\%$ and $B_v=0$ at $>99\%$. Also, mode coupling evidently contributes only a small fraction of the total difference between the \emph{sdbmc} and \emph{no sdbmc} results (the dotted curves are very close to the solid ones). This is consistent with theoretical expectations regarding the relative importance of mode coupling effects near the BAO scale for biased tracers (see Appendix~\ref{app:mc} and \cite{cs08,dcss10}). The right panels show that these statements are equally valid in Fourier space (see also Appendix~\ref{app:FourierMC}). Importantly, Table~\ref{tab:sdbmc-bestfit} and Fig.~\ref{fig:sdbmc-contours} show that the constraints on the mode coupling smearing scale $R_{\rm MC}$ are in excellent agreement with the theoretical value for \sigv\ reported in section~\ref{subsec:sims} for the HADES simulations.

\begin{table}[h]
\centering
\begin{tabular}{cccccc}
\hline\hline
$B_1$ & $B_v$ & $A_{\rm MC}$ & $R_{\rm MC}$  &  $\chi^2/{\rm dof}$ & $p$-value\\
&&& (\Mpch) && \\
\hline 
$0.60$ & $-14.3$ & $0.0077$ & $5.2$  & $118.7/98$ & $0.082$\\
\hline\hline
\end{tabular}
\caption{Best-fit values of the parameters $\{B_1,B_v,A_{\rm MC},R_{\rm MC}\}$ along with the $\chi^2$ per degree of freedom and corresponding $p$-value from the MCMC analysis of the HADES simulation measurements using the \emph{sdbmc} model. See Fig.~\ref{fig:sdbmc-contours} for the median and central $68\%$ confidence ranges of each parameter in the respective samples. This analysis used $b=1.95$ and $R_\ast=2.5\Mpch$ (see text for details).}
\label{tab:sdbmc-bestfit}
\end{table}

Interestingly, contrary to the naive expectation from peaks theory, the measurements prefer a significantly \emph{negative} best-fit value of the velocity bias coefficient $B_v$.\footnote{As we mentioned earlier, the intuition from peaks theory is usually built on the assumption of a narrow mass range. It is not difficult to show that, upon mass averaging the scale dependence of a typical peaks model, the result can indeed lead to inferring $B_v<0$. We will explore this using a detailed calculation in forthcoming work.}
We also see that, compared to the \emph{no sdbmc} model used by \citetalias{ps23}, the best fitting \emph{sdbmc} model predicts a distinctly narrower BAO peak in the monopole $\xiell{0}(s)$, which is also shifted to smaller $s$, and a larger amplitude of variations in the quadrupole $\xiell{2}(s)$. The hexadecapole $\xiell{4}(s)$, on the other hand, remains featureless, as for the \emph{no sdbmc} case. The analysis in Appendix~\ref{app:FourierMC} shows that these statements continue to hold when fitting the \emph{sdbmc} model to Fourier space measurements at higher redshift. Although the constraints on $R_{\rm MC}$ in this case (Table~\ref{tab:sdbmc-bestfit-kspace} and Fig.~\ref{fig:sdbmc-contours-kspace}) do not agree as well with the expected value \sigv\ in the MINERVA simulations, this is perhaps not surprising, given the limited range of validity of our approximation in Fourier space.

These results together underline the need for modelling the effects of scale-dependent bias at BAO scales, without which the inference of cosmological parameters could be substantially biased. E.g., since the \emph{no sdbmc} model has a broader BAO feature than the best fitting \emph{sdbmc} one, ignoring scale-dependent bias but allowing the large-scale bias $b$ to be free would at least lead to a significantly lower value of $b$ (driven by the data at $s\lesssim80\Mpch$ and $s\gtrsim110\Mpch$), in addition to a degradation of the quality of the fit (due to not being able to match the width of the BAO feature), in Fig.~\ref{fig:sdbmc-stats}. This would in turn potentially bias any constraints on parameters such as $f$ or $\sigma_8$ which have well-known degeneracies with $b$.


\subsection{Model variations}
Our models for the scale dependence of bias and mode coupling, although well-motivated as discussed in section~\ref{subsec:scaledepbias} and Appendix~\ref{app:mc}, are nevertheless somewhat \emph{ad hoc}. To assess the impact of various choices in the model setup and our analysis above, we also explored some variations which we briefly report below.

\subsubsection{Scale-dependent bias}
First, we considered the impact of allowing the smoothing scale $R_\ast$ to be free in the configuration space analysis, leading to 5 free parameters in the model. In this case, the value of $R_\ast$ is only weakly constrained with a best fit close to $\sim5\Mpch$, but still fully consistent with $2.5\Mpch$, and shows a strong degeneracy with $B_1$. The values of the other parameters are consistent with those reported in Table~\ref{tab:sdbmc-bestfit} and Fig.~\ref{fig:sdbmc-contours}, but with larger errors. In other words, the BAO-scale 2pcf measurements with DESI-like surveys are only weakly sensitive to the shape/scale of the smoothing filter defining scale-dependent bias. 

This has some consequences for our eventual goal of model-agnostic inference. The structure of \eqns{eq:B(k,mu)} and~\eqref{eq:Dellprop} shows that $R_\ast$ only appears in the combination $\e{-k^2(\sigv^2+R_\ast^2)}$. In a model-agnostic inference exercise, \sigv\ would be treated as a free parameter. Despite the fact that \sigv\ also appears in other factors such as $\e{-K^2\mu_k^2}$ in \eqn{eq:Dellprop}, its appearance along with $R_\ast^2$ implies that these parameters will likely be highly degenerate in a model-agnostic setting. This will need to be accounted for in any exercise aimed at constraining cosmological parameters like \sigv; we will discuss this further in future work.

Next, we also explored a more complex model of scale dependence, still within the Laplace-Gauss expansion framework, in which we allowed separate Gaussian smearing scales for the density and velocity bias factors, combined in a structure motivated by the shapes of the curves seen in fig.~1 of \cite{bds15}. However, the resulting 4-parameter model showed strong degeneracies in the parameters, without improving the quality of description of the clustering measurements.
We conclude that our 3-parameter description of scale-dependent bias in \eqns{eq:bLag-model} and~\eqref{eq:bvel-model} is sufficiently accurate and simple for our purposes.

\subsubsection{Mode coupling}
While the scale-dependence of density and velocity bias has been discussed widely in the literature, the effects of mode coupling for biased tracers have attracted considerably less attention. In part, this may be because, in Fourier space, the leading order effects of mode coupling are degenerate with those of scale dependent bias and are consequently treated as being part of `counter terms' in effective field theory (EFT) treatments \cite{chs12}.

Although our model separates out the effects of model coupling, we recall that the discussion in Appendix~\ref{app:mc} and our implementation in \eqref{eq:xiNL-sdbmc} (equivalently, equations~\ref{eq:xiellmc} and~\ref{eq:DellL}) implies that the mode coupling term depends on the scale-dependent bias of the tracers. To assess the impact of this feature of the model, we repeated our analysis by replacing the occurence of $B(k,\mu_k)$ in \eqn{eq:DellL} with $1+\beta\mu_k^2$. We found only statistically insignificant differences in the resulting parameter constraints and quality of fit. This indicates that, in practical terms, the constraints on the mode coupling piece are relatively insensitive to the details of scale-dependent bias. Together with the fact, noted earlier, that the mode coupling contribution is relatively small, this reassures us that our specific implementation of mode coupling is unlikely to produce biases in parameter recovery. We do advocate retaining the mode coupling contribution, given that its amplitude is constrained to be positive at $>95\%$ confidence (c.f. Fig.~\ref{fig:sdbmc-contours}). 



\section{Conclusion}
\label{sec:conclude}
Exploiting the BAO feature for model-agnostic cosmological inference is a challenging current problem for upcoming surveys such as DESI and Euclid. As identified by \citetalias{ps23}, one bottleneck in these efforts is the modelling of the scale dependence of galaxy bias, along with the effects of mode coupling, both of which are nonlinear effects that are known to modify the shape and location of the BAO feature. Modelling these effects in redshift space is critical in order to build robust model-agnostic frameworks for cosmological inference.

In this work, using simple and well-motivated models for the scale dependence of Lagrangian density and velocity bias and the effects of mode coupling, we have shown how to model the redshift-space multipoles of the 2-point correlation function (2pcf) accounting for these effects. Our model draws inspiration from the application of the Zel'dovich approximation to peaks theory (e.g., section~\ref{subsec:scaledepbias} and Appendix~\ref{app:mc}) and is agnostic to the details of the cosmological model, in that the predictions for the multipoles can be calculated for arbitrary linear power spectra. We have validated the model using $\Lambda$CDM simulations, finding that it can accurately describe the 2pcf multipoles $\xiell{\ell}(s)$ for $55\lesssim s/(\Mpch) \lesssim125$, i.e., near the BAO feature, for realistic tracer samples in configuration space. (In Appendix~\ref{app:FourierMC}, we also show that the model works well in Fourier space for $k\lesssim0.13\hMpc$, although this is less relevant for the model-agnostic applications we have in mind.)

Our model in \eqn{eq:xiNL-sdbmc} introduced a total of 5 new parameters: three of these $\{B_1,B_{\rm v},R_\ast\}$ describe the effects of scale-dependent bias, while the other two $\{A_{\rm MC},R_{\rm MC}\}$ additionally model the contribution of mode coupling. While this model allowed for the mode coupling contribution to be smeared by a scale $R_{\rm MC}$ that is not necessarily equal to the expected theoretical value of \sigv, our analysis of the HADES simulation in fact showed that it is perfectly acceptable to set $R_{\rm MC}=\sigv$. This would lead to a very useful reduction of the parameter space in future work, where \sigv\ will be varied as a free parameter.


A key advantage of our model is that it falls within the class of `Laplace-Gauss' expansions which were discussed by \citetalias{ps23} as being particularly useful in model-agnostic inference using the BAO feature in redshift space. A Laplace-Gauss expansion in Fourier space, i.e., an expression involving powers of $k^2$ multiplying terms like $\e{-k^2\sigma^2}$, combines naturally with the Zel'dovich smearing represented by the term $\e{-k^2\sigv^2}$ in the `propagator' term in \eqn{eq:Dellprop}. In forthcoming work, we will show how this can be leveraged to build Zel'dovich smearing approximations that account for both scale-dependent bias as well as mode coupling. Together with a set of well-motivated minimal basis functions for describing the linear theory 2pcf of real space matter fluctuations $\xi_{\rm lin}(r)$ \citep[e.g.,][]{nsz21b,lnps24,ps25}, this will help develop a complete framework for model-agnostic inference of cosmological parameters such as $f$, \sigv\ and the BAO scale.

\section*{Acknowledgments}
We are grateful to F. Nikakhtar for providing the multipole measurements in the HADES simulations.
The research of AP is supported by the Associates Scheme of ICTP, Trieste.  
AP and RKS thank ICTP, Trieste for hospitality during the summer of 2024 when part of this work was completed.
This work made extensive use of the open source computing packages NumPy \citep{vanderwalt-numpy},\footnote{\url{http://www.numpy.org}} SciPy \citep{scipy},\footnote{\url{http://www.scipy.org}} Matplotlib \citep{hunter07_matplotlib},\footnote{\url{https://matplotlib.org/}} and Jupyter Notebook.\footnote{\url{https://jupyter.org}} 

\bibliography{references}

\providecommand{\href}[2]{#2}\begingroup\raggedright\begin{thebibliography}{10}

\bibitem{eisenstein+05}
D.J.~{Eisenstein}, I.~{Zehavi}, D.W.~{Hogg}, R.~{Scoccimarro}, M.R.~{Blanton},
  R.C.~{Nichol} et~al., \emph{{Detection of the Baryon Acoustic Peak in the
  Large-Scale Correlation Function of SDSS Luminous Red Galaxies}},
  \href{https://doi.org/10.1086/466512}{\emph{\apj} {\bfseries 633} (2005) 560}
  [\href{https://arxiv.org/abs/astro-ph/0501171}{{\ttfamily
  astro-ph/0501171}}].

\bibitem{cole+05}
S.~{Cole}, W.J.~{Percival}, J.A.~{Peacock}, P.~{Norberg}, C.M.~{Baugh},
  C.S.~{Frenk} et~al., \emph{{The 2dF Galaxy Redshift Survey: power-spectrum
  analysis of the final data set and cosmological implications}},
  \href{https://doi.org/10.1111/j.1365-2966.2005.09318.x}{\emph{\mnras}
  {\bfseries 362} (2005) 505}
  [\href{https://arxiv.org/abs/astro-ph/0501174}{{\ttfamily
  astro-ph/0501174}}].

\bibitem{anderson+12}
L.~{Anderson}, E.~{Aubourg}, S.~{Bailey}, D.~{Bizyaev}, M.~{Blanton},
  A.S.~{Bolton} et~al., \emph{{The clustering of galaxies in the SDSS-III
  Baryon Oscillation Spectroscopic Survey: baryon acoustic oscillations in the
  Data Release 9 spectroscopic galaxy sample}},
  \href{https://doi.org/10.1111/j.1365-2966.2012.22066.x}{\emph{\mnras}
  {\bfseries 427} (2012) 3435}
  [\href{https://arxiv.org/abs/1203.6594}{{\ttfamily 1203.6594}}].

\bibitem{anderson+14}
L.~{Anderson}, E.~{Aubourg}, S.~{Bailey}, F.~{Beutler}, A.S.~{Bolton},
  J.~{Brinkmann} et~al., \emph{{The clustering of galaxies in the SDSS-III
  Baryon Oscillation Spectroscopic Survey: measuring D$_{A}$ and H at z = 0.57
  from the baryon acoustic peak in the Data Release 9 spectroscopic Galaxy
  sample}}, \href{https://doi.org/10.1093/mnras/stt2206}{\emph{\mnras}
  {\bfseries 439} (2014) 83} [\href{https://arxiv.org/abs/1303.4666}{{\ttfamily
  1303.4666}}].

\bibitem{alam+17}
S.~{Alam}, M.~{Ata}, S.~{Bailey}, F.~{Beutler}, D.~{Bizyaev}, J.A.~{Blazek}
  et~al., \emph{{The clustering of galaxies in the completed SDSS-III Baryon
  Oscillation Spectroscopic Survey: cosmological analysis of the DR12 galaxy
  sample}}, \href{https://doi.org/10.1093/mnras/stx721}{\emph{\mnras}
  {\bfseries 470} (2017) 2617}
  [\href{https://arxiv.org/abs/1607.03155}{{\ttfamily 1607.03155}}].

\bibitem{bottaro+24}
S.~{Bottaro}, E.~{Castorina}, M.~{Costa}, D.~{Redigolo} and E.~{Salvioni},
  \emph{{Unveiling Dark Forces with Measurements of the Large Scale Structure
  of the Universe}},
  \href{https://doi.org/10.1103/PhysRevLett.132.201002}{\emph{\prl} {\bfseries
  132} (2024) 201002} [\href{https://arxiv.org/abs/2309.11496}{{\ttfamily
  2309.11496}}].

\bibitem{mhc24}
R.~{Mauland}, H.A.~{Winther} and C.-Z.~{Ruan}, \emph{{Sesame: A power spectrum
  emulator pipeline for beyond-{\ensuremath{\Lambda}}CDM models}},
  \href{https://doi.org/10.1051/0004-6361/202347892}{\emph{\aap} {\bfseries
  685} (2024) A156} [\href{https://arxiv.org/abs/2309.13295}{{\ttfamily
  2309.13295}}].

\bibitem{pmp24}
S.~{Paradiso}, G.~{McGee} and W.J.~{Percival}, \emph{{Evaluating extensions to
  LCDM: an application of Bayesian model averaging and selection}},
  \href{https://doi.org/10.48550/arXiv.2403.02120}{\emph{arXiv e-prints} (2024)
  arXiv:2403.02120} [\href{https://arxiv.org/abs/2403.02120}{{\ttfamily
  2403.02120}}].

\bibitem{cuesta+16}
A.J.~{Cuesta}, M.~{Vargas-Maga{\~n}a}, F.~{Beutler}, A.S.~{Bolton},
  J.R.~{Brownstein}, D.J.~{Eisenstein} et~al., \emph{{The clustering of
  galaxies in the SDSS-III Baryon Oscillation Spectroscopic Survey: baryon
  acoustic oscillations in the correlation function of LOWZ and CMASS galaxies
  in Data Release 12}},
  \href{https://doi.org/10.1093/mnras/stw066}{\emph{\mnras} {\bfseries 457}
  (2016) 1770} [\href{https://arxiv.org/abs/1509.06371}{{\ttfamily
  1509.06371}}].

\bibitem{beutler+17}
F.~{Beutler}, H.-J.~{Seo}, S.~{Saito}, C.-H.~{Chuang}, A.J.~{Cuesta},
  D.J.~{Eisenstein} et~al., \emph{{The clustering of galaxies in the completed
  SDSS-III Baryon Oscillation Spectroscopic Survey: anisotropic galaxy
  clustering in Fourier space}},
  \href{https://doi.org/10.1093/mnras/stw3298}{\emph{\mnras} {\bfseries 466}
  (2017) 2242} [\href{https://arxiv.org/abs/1607.03150}{{\ttfamily
  1607.03150}}].

\bibitem{blomqvist+19}
M.~{Blomqvist}, H.~{du Mas des Bourboux}, N.G.~{Busca}, V.~{de Sainte Agathe},
  J.~{Rich}, C.~{Balland} et~al., \emph{{Baryon acoustic oscillations from the
  cross-correlation of Ly{\ensuremath{\alpha}} absorption and quasars in eBOSS
  DR14}}, \href{https://doi.org/10.1051/0004-6361/201935641}{\emph{\aap}
  {\bfseries 629} (2019) A86}
  [\href{https://arxiv.org/abs/1904.03430}{{\ttfamily 1904.03430}}].

\bibitem{dumasdesbourbouz+20}
H.~{du Mas des Bourboux}, J.~{Rich}, A.~{Font-Ribera}, V.~{de Sainte Agathe},
  J.~{Farr}, T.~{Etourneau} et~al., \emph{{The Completed SDSS-IV Extended
  Baryon Oscillation Spectroscopic Survey: Baryon Acoustic Oscillations with
  Ly{\ensuremath{\alpha}} Forests}},
  \href{https://doi.org/10.3847/1538-4357/abb085}{\emph{\apj} {\bfseries 901}
  (2020) 153} [\href{https://arxiv.org/abs/2007.08995}{{\ttfamily
  2007.08995}}].

\bibitem{gil-marin+20}
H.~{Gil-Mar{\'\i}n}, J.E.~{Bautista}, R.~{Paviot}, M.~{Vargas-Maga{\~n}a},
  S.~{de la Torre}, S.~{Fromenteau} et~al., \emph{{The Completed SDSS-IV
  extended Baryon Oscillation Spectroscopic Survey: measurement of the BAO and
  growth rate of structure of the luminous red galaxy sample from the
  anisotropic power spectrum between redshifts 0.6 and 1.0}},
  \href{https://doi.org/10.1093/mnras/staa2455}{\emph{\mnras} {\bfseries 498}
  (2020) 2492} [\href{https://arxiv.org/abs/2007.08994}{{\ttfamily
  2007.08994}}].

\bibitem{extractor2020}
E.~{Noda}, M.~{Peloso} and M.~{Pietroni}, \emph{{Extracting the BAO scale from
  BOSS DR12 dataset}},
  \href{https://doi.org/10.1016/j.dark.2020.100579}{\emph{Physics of the Dark
  Universe} {\bfseries 29} (2020) 100579}
  [\href{https://arxiv.org/abs/1901.06854}{{\ttfamily 1901.06854}}].

\bibitem{abbott+22}
T.M.C.~{Abbott}, M.~{Aguena}, S.~{Allam}, A.~{Amon}, F.~{Andrade-Oliveira},
  J.~{Asorey} et~al., \emph{{Dark Energy Survey Year 3 results: A 2.7\%
  measurement of baryon acoustic oscillation distance scale at redshift
  0.835}}, \href{https://doi.org/10.1103/PhysRevD.105.043512}{\emph{\prd}
  {\bfseries 105} (2022) 043512}
  [\href{https://arxiv.org/abs/2107.04646}{{\ttfamily 2107.04646}}].

\bibitem{bst24}
I.~{Babi{\'c}}, F.~{Schmidt} and B.~{Tucci}, \emph{{Straightening the Ruler:
  Field-Level Inference of the BAO Scale with LEFTfield}},
  \href{https://doi.org/10.48550/arXiv.2407.01524}{\emph{arXiv e-prints} (2024)
  arXiv:2407.01524} [\href{https://arxiv.org/abs/2407.01524}{{\ttfamily
  2407.01524}}].

\bibitem{bst25}
I.~{Babi{\'c}}, F.~{Schmidt} and B.~{Tucci}, \emph{{Forward vs Backward:
  Improving BAO Constraints with Field-Level Inference}},
  \href{https://doi.org/10.48550/arXiv.2505.13588}{\emph{arXiv e-prints} (2025)
  arXiv:2505.13588} [\href{https://arxiv.org/abs/2505.13588}{{\ttfamily
  2505.13588}}].

\bibitem{LP2016}
S.~{Anselmi}, G.D.~{Starkman} and R.K.~{Sheth}, \emph{{Beating non-linearities:
  improving the baryon acoustic oscillations with the linear point}},
  \href{https://doi.org/10.1093/mnras/stv2436}{\emph{\mnras} {\bfseries 455}
  (2016) 2474} [\href{https://arxiv.org/abs/1508.01170}{{\ttfamily
  1508.01170}}].

\bibitem{LPmocks}
S.~{Anselmi}, P.-S.~{Corasaniti}, G.D.~{Starkman}, R.K.~{Sheth} and
  I.~{Zehavi}, \emph{{Linear point standard ruler for galaxy survey data:
  Validation with mock catalogs}},
  \href{https://doi.org/10.1103/PhysRevD.98.023527}{\emph{\prd} {\bfseries 98}
  (2018) 023527} [\href{https://arxiv.org/abs/1711.09063}{{\ttfamily
  1711.09063}}].

\bibitem{LPboss}
S.~{Anselmi}, G.D.~{Starkman}, P.-S.~{Corasaniti}, R.K.~{Sheth} and
  I.~{Zehavi}, \emph{{Galaxy Correlation Functions Provide a More Robust
  Cosmological Standard Ruler}},
  \href{https://doi.org/10.1103/PhysRevLett.121.021302}{\emph{\prl} {\bfseries
  121} (2018) 021302} [\href{https://arxiv.org/abs/1703.01275}{{\ttfamily
  1703.01275}}].

\bibitem{nsz21a}
F.~{Nikakhtar}, R.K.~{Sheth} and I.~{Zehavi}, \emph{{Laguerre reconstruction of
  the correlation function on baryon acoustic oscillation scales}},
  \href{https://doi.org/10.1103/PhysRevD.104.043530}{\emph{\prd} {\bfseries
  104} (2021) 043530} [\href{https://arxiv.org/abs/2101.08376}{{\ttfamily
  2101.08376}}].

\bibitem{nsz21b}
F.~{Nikakhtar}, R.K.~{Sheth} and I.~{Zehavi}, \emph{{Laguerre reconstruction of
  the BAO feature in halo-based mock galaxy catalogues}},
  \href{https://doi.org/10.1103/PhysRevD.104.063504}{\emph{\prd} {\bfseries
  104} (2021) 063504} [\href{https://arxiv.org/abs/2107.12537}{{\ttfamily
  2107.12537}}].

\bibitem{ps22}
A.~{Paranjape} and R.K.~{Sheth}, \emph{{Bayesian evidence comparison for
  distance scale estimates}},
  \href{https://doi.org/10.1093/mnras/stac2984}{\emph{\mnras} {\bfseries 517}
  (2022) 4696} [\href{https://arxiv.org/abs/2209.00668}{{\ttfamily
  2209.00668}}].

\bibitem{ps23}
A.~{Paranjape} and R.K.~{Sheth}, \emph{{Model-agnostic cosmological constraints
  from the baryon acoustic oscillation feature in redshift space}},
  \href{https://doi.org/10.1093/mnras/stad2741}{\emph{\mnras} (2023) }
  [\href{https://arxiv.org/abs/2304.09198}{{\ttfamily 2304.09198}}].

\bibitem{bharadwaj96}
S.~{Bharadwaj}, \emph{{The Evolution of Correlation Functions in the Zeldovich
  Approximation and Its Implications for the Validity of Perturbation Theory}},
  \href{https://doi.org/10.1086/178036}{\emph{\apj} {\bfseries 472} (1996) 1}
  [\href{https://arxiv.org/abs/astro-ph/9606121}{{\ttfamily
  astro-ph/9606121}}].

\bibitem{cs06b}
M.~{Crocce} and R.~{Scoccimarro}, \emph{{Memory of initial conditions in
  gravitational clustering}},
  \href{https://doi.org/10.1103/PhysRevD.73.063520}{\emph{\prd} {\bfseries 73}
  (2006) 063520} [\href{https://arxiv.org/abs/astro-ph/0509419}{{\ttfamily
  astro-ph/0509419}}].

\bibitem{DESI}
{DESI Collaboration}, A.~{Aghamousa}, J.~{Aguilar}, S.~{Ahlen}, S.~{Alam},
  L.E.~{Allen} et~al., \emph{{The DESI Experiment Part I: Science,Targeting,
  and Survey Design}}, {\emph{arXiv e-prints} (2016) arXiv:1611.00036}
  [\href{https://arxiv.org/abs/1611.00036}{{\ttfamily 1611.00036}}].

\bibitem{DESI2024-cosmo}
{DESI Collaboration}, A.G.~{Adame}, J.~{Aguilar}, S.~{Ahlen}, S.~{Alam},
  D.M.~{Alexander} et~al., \emph{{DESI 2024 VI: Cosmological Constraints from
  the Measurements of Baryon Acoustic Oscillations}},
  \href{https://doi.org/10.48550/arXiv.2404.03002}{\emph{arXiv e-prints} (2024)
  arXiv:2404.03002} [\href{https://arxiv.org/abs/2404.03002}{{\ttfamily
  2404.03002}}].

\bibitem{dcss10}
V.~{Desjacques}, M.~{Crocce}, R.~{Scoccimarro} and R.K.~{Sheth},
  \emph{{Modeling scale-dependent bias on the baryonic acoustic scale with the
  statistics of peaks of Gaussian random fields}},
  \href{https://doi.org/10.1103/PhysRevD.82.103529}{\emph{\prd} {\bfseries 82}
  (2010) 103529} [\href{https://arxiv.org/abs/1009.3449}{{\ttfamily
  1009.3449}}].

\bibitem{bkBAO}
S.~{Gaines}, F.~{Nikakhtar}, N.~{Padmanabhan} and R.K.~{Sheth},
  \emph{{Leveraging protohalos and scale-dependent bias to calibrate the BAO
  scale in real space}},
  \href{https://doi.org/10.1103/PhysRevD.110.103511}{\emph{\prd} {\bfseries
  110} (2024) 103511} [\href{https://arxiv.org/abs/2408.00072}{{\ttfamily
  2408.00072}}].

\bibitem{mpp24}
M.~{Marinucci}, K.~{Pardede} and M.~{Pietroni}, \emph{{Bootstrapping Lagrangian
  Perturbation Theory for the Large Scale Structure}},
  \href{https://doi.org/10.48550/arXiv.2405.08413}{\emph{arXiv e-prints} (2024)
  arXiv:2405.08413} [\href{https://arxiv.org/abs/2405.08413}{{\ttfamily
  2405.08413}}].

\bibitem{cs06a}
M.~{Crocce} and R.~{Scoccimarro}, \emph{{Renormalized cosmological perturbation
  theory}}, \href{https://doi.org/10.1103/PhysRevD.73.063519}{\emph{\prd}
  {\bfseries 73} (2006) 063519}
  [\href{https://arxiv.org/abs/astro-ph/0509418}{{\ttfamily
  astro-ph/0509418}}].

\bibitem{cs08}
M.~{Crocce} and R.~{Scoccimarro}, \emph{{Nonlinear evolution of baryon acoustic
  oscillations}}, \href{https://doi.org/10.1103/PhysRevD.77.023533}{\emph{\prd}
  {\bfseries 77} (2008) 023533}
  [\href{https://arxiv.org/abs/0704.2783}{{\ttfamily 0704.2783}}].

\bibitem{ZeldovichRSD}
A.N.~{Taylor} and A.J.S.~{Hamilton}, \emph{{Non-linear cosmological power
  spectra in real and redshift space}},
  \href{https://doi.org/10.1093/mnras/282.3.767}{\emph{\mnras} {\bfseries 282}
  (1996) 767} [\href{https://arxiv.org/abs/astro-ph/9604020}{{\ttfamily
  astro-ph/9604020}}].

\bibitem{ds10}
V.~{Desjacques} and R.K.~{Sheth}, \emph{{Redshift space correlations and
  scale-dependent stochastic biasing of density peaks}},
  \href{https://doi.org/10.1103/PhysRevD.81.023526}{\emph{\prd} {\bfseries 81}
  (2010) 023526} [\href{https://arxiv.org/abs/0909.4544}{{\ttfamily
  0909.4544}}].

\bibitem{ppvv15}
M.~{Peloso}, M.~{Pietroni}, M.~{Viel} and F.~{Villaescusa-Navarro}, \emph{{The
  effect of massive neutrinos on the BAO peak}},
  \href{https://doi.org/10.1088/1475-7516/2015/07/001}{\emph{\jcap} {\bfseries
  2015} (2015) 001} [\href{https://arxiv.org/abs/1505.07477}{{\ttfamily
  1505.07477}}].

\bibitem{kaiser87}
N.~{Kaiser}, \emph{{Clustering in real space and in redshift space}},
  \href{https://doi.org/10.1093/mnras/227.1.1}{\emph{\mnras} {\bfseries 227}
  (1987) 1}.

\bibitem{bds15}
T.~{Baldauf}, V.~{Desjacques} and U.~{Seljak}, \emph{{Velocity bias in the
  distribution of dark matter halos}},
  \href{https://doi.org/10.1103/PhysRevD.92.123507}{\emph{\prd} {\bfseries 92}
  (2015) 123507} [\href{https://arxiv.org/abs/1405.5885}{{\ttfamily
  1405.5885}}].

\bibitem{phs18}
A.~{Paranjape}, O.~{Hahn} and R.K.~{Sheth}, \emph{{Halo assembly bias and the
  tidal anisotropy of the local halo environment}},
  \href{https://doi.org/10.1093/mnras/sty496}{\emph{\mnras} {\bfseries 476}
  (2018) 3631} [\href{https://arxiv.org/abs/1706.09906}{{\ttfamily
  1706.09906}}].

\bibitem{smt01}
R.K.~{Sheth}, H.J.~{Mo} and G.~{Tormen}, \emph{{Ellipsoidal collapse and an
  improved model for the number and spatial distribution of dark matter
  haloes}},
  \href{https://doi.org/10.1046/j.1365-8711.2001.04006.x}{\emph{\mnras}
  {\bfseries 323} (2001) 1}
  [\href{https://arxiv.org/abs/arXiv:astro-ph/9907024}{{\ttfamily
  arXiv:astro-ph/9907024}}].

\bibitem{st02}
R.K.~{Sheth} and G.~{Tormen}, \emph{{An excursion set model of hierarchical
  clustering: ellipsoidal collapse and the moving barrier}},
  \href{https://doi.org/10.1046/j.1365-8711.2002.04950.x}{\emph{\mnras}
  {\bfseries 329} (2002) 61}
  [\href{https://arxiv.org/abs/astro-ph/0105113}{{\ttfamily
  astro-ph/0105113}}].

\bibitem{cphs17}
E.~{Castorina}, A.~{Paranjape}, O.~{Hahn} and R.K.~{Sheth}, \emph{{Excursion
  set peaks: the role of shear}}, {\emph{ArXiv e-prints} (2016) }
  [\href{https://arxiv.org/abs/1611.03619}{{\ttfamily 1611.03619}}].

\bibitem{hades}
F.~{Villaescusa-Navarro}, A.~{Banerjee}, N.~{Dalal}, E.~{Castorina},
  R.~{Scoccimarro}, R.~{Angulo} et~al., \emph{{The Imprint of Neutrinos on
  Clustering in Redshift Space}},
  \href{https://doi.org/10.3847/1538-4357/aac6bf}{\emph{\apj} {\bfseries 861}
  (2018) 53} [\href{https://arxiv.org/abs/1708.01154}{{\ttfamily 1708.01154}}].

\bibitem{nikakhtar+23}
F.~{Nikakhtar}, N.~{Padmanabhan}, B.~{L{\'e}vy}, R.K.~{Sheth} and
  R.~{Mohayaee}, \emph{{Optimal transport reconstruction of biased tracers in
  redshift space}},
  \href{https://doi.org/10.1103/PhysRevD.108.083534}{\emph{\prd} {\bfseries
  108} (2023) 083534} [\href{https://arxiv.org/abs/2307.03671}{{\ttfamily
  2307.03671}}].

\bibitem{grieb2016}
J.N.~{Grieb}, A.G.~{S{\'a}nchez}, S.~{Salazar-Albornoz} and C.~{Dalla Vecchia},
  \emph{{Gaussian covariance matrices for anisotropic galaxy clustering
  measurements}}, \href{https://doi.org/10.1093/mnras/stw065}{\emph{\mnras}
  {\bfseries 457} (2016) 1577}
  [\href{https://arxiv.org/abs/1509.04293}{{\ttfamily 1509.04293}}].

\bibitem{BOSSDR12-FinalData}
S.~{Alam}, F.D.~{Albareti}, C.~{Allende Prieto}, F.~{Anders}, S.F.~{Anderson},
  T.~{Anderton} et~al., \emph{{The Eleventh and Twelfth Data Releases of the
  Sloan Digital Sky Survey: Final Data from SDSS-III}},
  \href{https://doi.org/10.1088/0067-0049/219/1/12}{\emph{\apjs} {\bfseries
  219} (2015) 12} [\href{https://arxiv.org/abs/1501.00963}{{\ttfamily
  1501.00963}}].

\bibitem{tl19-cobaya}
J.~{Torrado} and A.~{Lewis}, ``{Cobaya: Bayesian analysis in cosmology}.''
  Astrophysics Source Code Library, record ascl:1910.019, Oct., 2019.

\bibitem{tl21-cobaya}
J.~{Torrado} and A.~{Lewis}, \emph{{Cobaya: code for Bayesian analysis of
  hierarchical physical models}},
  \href{https://doi.org/10.1088/1475-7516/2021/05/057}{\emph{\jcap} {\bfseries
  2021} (2021) 057} [\href{https://arxiv.org/abs/2005.05290}{{\ttfamily
  2005.05290}}].

\bibitem{lewis19}
A.~{Lewis}, \emph{{GetDist: a Python package for analysing Monte Carlo
  samples}}, \href{https://doi.org/10.48550/arXiv.1910.13970}{\emph{arXiv
  e-prints} (2019) arXiv:1910.13970}
  [\href{https://arxiv.org/abs/1910.13970}{{\ttfamily 1910.13970}}].

\bibitem{class-I}
J.~{Lesgourgues}, \emph{{The Cosmic Linear Anisotropy Solving System (CLASS) I:
  Overview}}, {\emph{arXiv e-prints} (2011) arXiv:1104.2932}
  [\href{https://arxiv.org/abs/1104.2932}{{\ttfamily 1104.2932}}].

\bibitem{class-II}
D.~{Blas}, J.~{Lesgourgues} and T.~{Tram}, \emph{{The Cosmic Linear Anisotropy
  Solving System (CLASS). Part II: Approximation schemes}},
  \href{https://doi.org/10.1088/1475-7516/2011/07/034}{\emph{\jcap} {\bfseries
  2011} (2011) 034} [\href{https://arxiv.org/abs/1104.2933}{{\ttfamily
  1104.2933}}].

\bibitem{Tinker10}
J.L.~{Tinker}, B.E.~{Robertson}, A.V.~{Kravtsov}, A.~{Klypin}, M.S.~{Warren},
  G.~{Yepes} et~al., \emph{{The Large-scale Bias of Dark Matter Halos:
  Numerical Calibration and Model Tests}},
  \href{https://doi.org/10.1088/0004-637X/724/2/878}{\emph{\apj} {\bfseries
  724} (2010) 878} [\href{https://arxiv.org/abs/1001.3162}{{\ttfamily
  1001.3162}}].

\bibitem{Tinker08}
J.~{Tinker}, A.V.~{Kravtsov}, A.~{Klypin}, K.~{Abazajian}, M.~{Warren},
  G.~{Yepes} et~al., \emph{{Toward a Halo Mass Function for Precision
  Cosmology: The Limits of Universality}},
  \href{https://doi.org/10.1086/591439}{\emph{\apj} {\bfseries 688} (2008) 709}
  [\href{https://arxiv.org/abs/0803.2706}{{\ttfamily 0803.2706}}].

\bibitem{chs12}
J.J.M.~{Carrasco}, M.P.~{Hertzberg} and L.~{Senatore}, \emph{{The effective
  field theory of cosmological large scale structures}},
  \href{https://doi.org/10.1007/JHEP09(2012)082}{\emph{Journal of High Energy
  Physics} {\bfseries 2012} (2012) 82}
  [\href{https://arxiv.org/abs/1206.2926}{{\ttfamily 1206.2926}}].

\bibitem{lnps24}
J.J.~{Lee}, F.~{Nikakhtar}, A.~{Paranjape} and R.K.~{Sheth},
  \emph{{Eigen-decomposition of Covariance matrices: An application to the BAO
  Linear Point}}, \href{https://doi.org/10.48550/arXiv.2407.04692}{\emph{arXiv
  e-prints} (2024) arXiv:2407.04692}
  [\href{https://arxiv.org/abs/2407.04692}{{\ttfamily 2407.04692}}].

\bibitem{ps25}
A.~{Paranjape} and R.K.~{Sheth}, \emph{{Model-agnostic basis functions for the
  2-point correlation function of dark matter in linear theory}},
  \href{https://doi.org/10.48550/arXiv.2410.21374}{\emph{arXiv e-prints} (2024)
  arXiv:2410.21374} [\href{https://arxiv.org/abs/2410.21374}{{\ttfamily
  2410.21374}}].

\bibitem{vanderwalt-numpy}
S.~{Van Der Walt}, S.C.~{Colbert} and G.~{Varoquaux}, \emph{{The NumPy array: a
  structure for efficient numerical computation}}, {\emph{ArXiv e-prints}
  (2011) } [\href{https://arxiv.org/abs/1102.1523}{{\ttfamily 1102.1523}}].

\bibitem{scipy}
P.~Virtanen, R.~Gommers, T.E.~Oliphant, M.~Haberland, T.~Reddy, D.~Cournapeau
  et~al., \emph{{{SciPy} 1.0: Fundamental Algorithms for Scientific Computing
  in Python}}, \href{https://doi.org/10.1038/s41592-019-0686-2}{\emph{Nature
  Methods} {\bfseries 17} (2020) 261}.

\bibitem{hunter07_matplotlib}
J.D.~Hunter, \emph{Matplotlib: A 2d graphics environment},
  \href{https://doi.org/10.1109/MCSE.2007.55}{\emph{Computing In Science \&
  Engineering} {\bfseries 9} (2007) 90}.

\bibitem{peaksBAO}
T.~{Baldauf} and V.~{Desjacques}, \emph{{Phenomenology of baryon acoustic
  oscillation evolution from Lagrangian to Eulerian space}},
  \href{https://doi.org/10.1103/PhysRevD.95.043535}{\emph{\prd} {\bfseries 95}
  (2017) 043535} [\href{https://arxiv.org/abs/1612.04521}{{\ttfamily
  1612.04521}}].

\bibitem{iPTpeaks}
T.~{Matsubara}, \emph{{Velocity bias and the nonlinear perturbation theory of
  peaks}}, \href{https://doi.org/10.1103/PhysRevD.100.083504}{\emph{\prd}
  {\bfseries 100} (2019) 083504}
  [\href{https://arxiv.org/abs/1907.13251}{{\ttfamily 1907.13251}}].

\bibitem{peaksRSD}
V.~{Desjacques} and R.K.~{Sheth}, \emph{{Redshift space correlations and
  scale-dependent stochastic biasing of density peaks}},
  \href{https://doi.org/10.1103/PhysRevD.81.023526}{\emph{\prd} {\bfseries 81}
  (2010) 023526} [\href{https://arxiv.org/abs/0909.4544}{{\ttfamily
  0909.4544}}].

\end{thebibliography}\endgroup
\appendix

\section{Fourier space results}
\label{app:FourierMC}
Here, we test whether the \emph{sdbmc} model described in the main text can also describe redshift space non-linearities in Fourier space, by comparing the Fourier space counterparts $\Dellsq{\ell}(k)$ of the multipoles $\xiell{\ell}(s)$ with corresponding  measurements at $z=0.57$ in the simulations of \cite{grieb2016}. 

While $\Dellpropsq{\ell}(k)$ in \eqn{eq:Dellprop} can be identified as the Fourier space multipole of the `propagator' term, the structure of \eqn{eq:xiellmc} shows that $\DellLsq{\ell}(k)$ in \eqn{eq:DellL} is not equal to the Fourier multipole $\Dellmcsq{\ell}(k)$ of the mode coupling term, rather, it is the multipole of the linear theory anisotropic power spectrum (modified by the scale-dependent bias terms). Instead, we have
\begin{align}
\Dellmcsq{\ell}(k) &= \frac{2k^3}{\pi}\,(-i)^\ell\int_0^\infty\der s\,s^2\,j_\ell(ks)\,\xiellmc{\ell}(s)
\end{align}
%
To simplify this, we define the integral $\Cal{I}_\ell(k,k^\prime)$ as
\beq
\Cal{I}_\ell(k,k^\prime) \equiv \int_0^\infty\der s\,s^3\,j_\ell(ks)\,j_{\ell+1}(k^\prime s)\,,
\eeq
which leads to
\begin{align}
\Dellmcsq{\ell}(k) &= \frac{2k^3}{\pi}\,(-i)^\ell\int_0^\infty\der s\,s^2\,j_\ell(ks)\,\xiellmc{\ell}(s)\notag\\
&= A_{\rm MC}\,\frac{2k^3}{\pi}\,(-i^2)^\ell \int\der\ln k^\prime \DellLsq{\ell}(k^\prime) \int_0^\infty\der s\,s^2\,j_\ell(ks)\left[\ell\,j_\ell(k^\prime s) - k^\prime s\,j_{\ell+1}(k^\prime s)\right] \notag\\
&= A_{\rm MC}\,\frac{2k^3}{\pi} \int_0^\infty\der k^\prime\,\DellLsq{\ell}(k^\prime) \left[\frac{\ell\,\pi}{2k^3}\dir(k-k^\prime) - \Cal{I}_\ell(k,k^\prime)\right] \notag\\
&= A_{\rm MC} \left[\ell\,\DellLsq{\ell}(k) - \frac{2k^3}{\pi}\int_0^\infty\der k^\prime\,\DellLsq{\ell}(k^\prime)\,\Cal{I}_\ell(k,k^\prime) \right]\,,
\label{eq:DellMC-simple}
\end{align}
where we used the closure identity
\beq
\int_0^\infty\der x\,x^2\,j_\ell(a x)\,j_\ell(b x) = \frac{\pi}{2a^2}\,\dir(a-b)\,.
\label{eq:sphbessel-closure}
\eeq
To proceed further, we use the recursion relation
\beq
j_{\ell+1}(x) = (2\ell+1)\,j_\ell(x)/x - j_{\ell-1}(x)\,,
\label{eq:sphbessel-recur}
\eeq
along with the closure relation \eqref{eq:sphbessel-closure} to derive the recursion relation
\beq
\Cal{I}_{\ell+1}(k,k^\prime) = \frac{(2\ell+1)\pi}{2k^3}\,\dir(k-k^\prime) - \Cal{I}_{\ell-1}(k^\prime,k)\,.
\label{eq:I_ell-recur}
\eeq
Note the exchange of arguments in the last term. This recursion can be used if we have an explicit expression for $\Cal{I}_0(k,k^\prime)$, which can be derived as follows.
\begin{align}
\Cal{I}_0(k,k^\prime) &= \int_0^\infty\der s\,s^3\,\frac{\sin(ks)}{ks}\,\frac{1}{k^{\prime2}s^2}\left[\sin(k^\prime s) - k^\prime s\,\cos(k^\prime s)\right]\notag\\
&= \int_0^\infty\der s\,\frac{\sin(ks)}{kk^{\prime2}}\,\left[\sin(k^\prime s) - k^\prime s\,\cos(k^\prime s)\right] \notag\\
\therefore \Cal{I}_0(k,k^\prime) &= \frac{\pi}{2kk^\prime}\left[\frac{1}{k^\prime}\left(\dir(k-k^\prime) - \dir(k+k^\prime)\right) - \frac{\p}{\p k^\prime}\left(\dir(k-k^\prime) - \dir(k+k^\prime)\right)\right]\,,
\label{eq:I_0-def}
\end{align}
where we used the Dirac delta expression $\int_{-\infty}^\infty\der x\,\e{iax} = 2\pi\,\dir(a)$ to write
\begin{align}
\int_0^\infty\der x\,\sin(ax)\,\sin(bx) &= \frac{\pi}{2}\left[\dir(a-b)-\dir(a+b)\right]\,,\\
\int_0^\infty\der x\,x\,\sin(ax)\,\cos(bx) &= \frac{\pi}{2}\,\frac{\p}{\p b}\left[\dir(a-b)-\dir(a+b)\right] \,.
\end{align}
We also note that the terms involving $\dir(k+k^\prime)$ in \eqn{eq:I_0-def} will never contribute, since $k>0$ and the integral over $k^\prime$ is also always restricted to positive values in \eqn{eq:DellMC-simple}, so that we have
\beq
\frac{2k^3}{\pi}\int_0^\infty\der k^\prime\,F(k^\prime)\,\Cal{I}_0(k,k^\prime) = k\,\p_k F\,,
\eeq
for any function $F(k)$.

\begin{figure*}[h]
\centering
\includegraphics[width=0.48\textwidth]{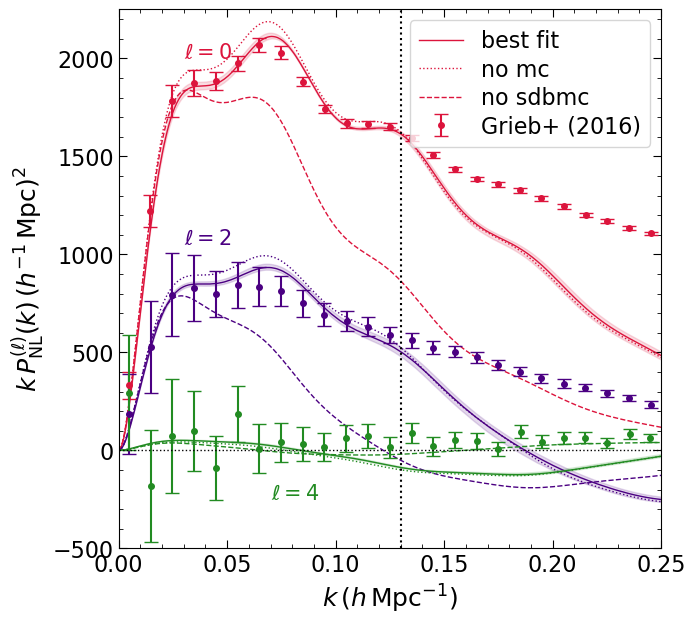}
\includegraphics[width=0.46\textwidth]{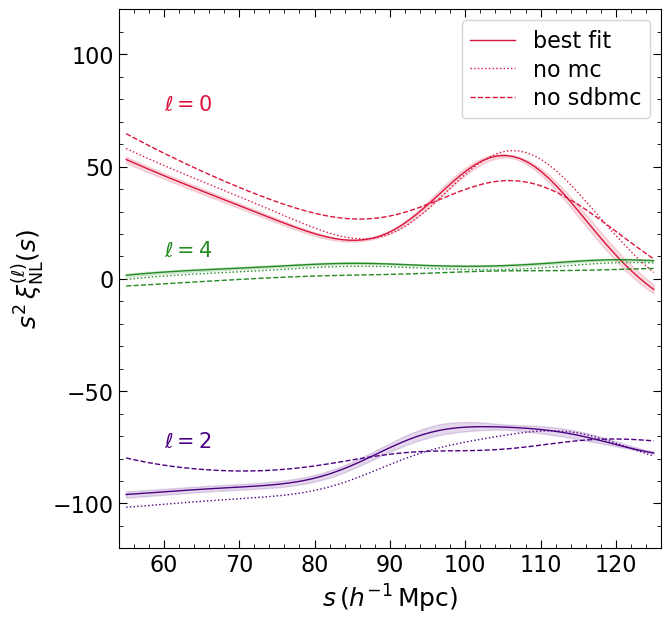}
\caption{\emph{(Left panel):} Comparison of the Fourier-space measurements from \cite{grieb2016} (points with error bars) with the best fitting \emph{sdbmc} model (equation~\ref{eq:Delta(ell)2-sdbmc-def}; solid curves with error bands). For comparison, the dashed curves show the \emph{no sdbmc} model (equation~\ref{eq:Delta(ell)2-def}) and the dotted curves show the result of setting $A_{\rm MC}=0$ in the best-fit  \emph{sdbmc} model (c.f. Fig.~\ref{fig:sdbmc-stats}). Vertical dotted line indicates the threshold $k_{\rm max}=0.13\,\hMpc$; the likelihood was evaluated using data with $k<k_{\rm max}$. \emph{(Right panel):} Configuration space results corresponding the the Fourier-space ones from the left panel, formatted identically.
This analysis used $b=2.005$ and $R_\ast=2.5\Mpch$, with cosmological parameters set to the ones used by \cite{grieb2016} (see text for details). 
}
\label{fig:sdbmc-stats-kspace}
\end{figure*}

A straightforward calculation using \eqns{eq:I_ell-recur} and \eqref{eq:I_0-def} in \eqn{eq:DellMC-simple} now gives
\begin{align}
\Dellmcsq{0}(k) &= -A_{\rm MC}\, k\,\p_k\DellLsq{0}(k)\,, \label{eq:D0MC}\\
\Dellmcsq{2}(k) &= -A_{\rm MC}\, k\,\p_k\DellLsq{2}(k)\,, \label{eq:D2MC}\\
\Dellmcsq{4}(k) &= A_{\rm MC} \left(4\DellLsq{4}(k)+ k\,\p_k\DellLsq{4}(k) \right)\,. \label{eq:D4MC}
\end{align}
These expressions are useful in comparing this model to the $k$-space multipole measurements of \cite{grieb2016}. 
The \emph{sdbmc} model including effects of scale-dependent bias and mode coupling is then
\begin{align}
\Dellsq{\ell}(k) &= \Dellpropsq{\ell}(k) + \Dellmcsq{\ell}(k)\,,
\label{eq:Delta(ell)2-sdbmc-def}
\end{align}
with $\Dellpropsq{\ell}(k)$ given by \eqn{eq:Dellprop} and $\Dellmcsq{\ell}(k)$ by \eqns{eq:D0MC}-\eqref{eq:D4MC}. 

\begin{figure}[h]
\centering
\includegraphics[width=0.65\textwidth]{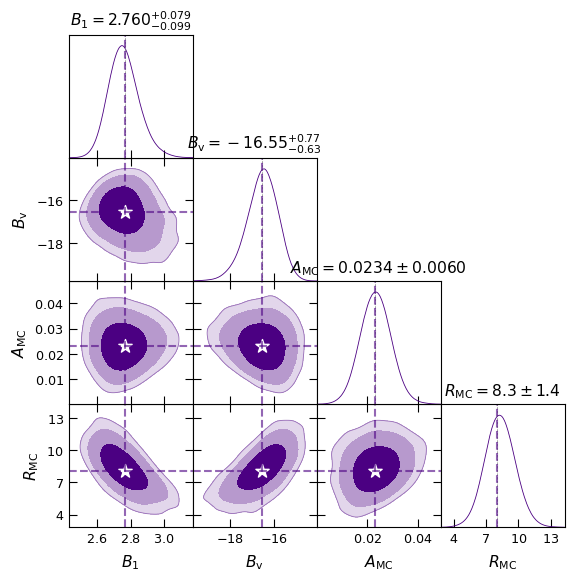}
\caption{Constraints on \emph{sdbmc} model parameters using the Fourier space measurements of \cite{grieb2016}. The contours show the $68\%$, $95\%$ and $99\%$ confidence regions, and the dashed lines intersecting at the white stars show the best-fit parameter combination.
This analysis used $b=2.005$ and $R_\ast=2.5\Mpch$, with cosmological parameters set to the ones used by \cite{grieb2016} (see text for details). 
}
\label{fig:sdbmc-contours-kspace}
\end{figure}

The results of varying the $4$ parameters $\{B_1,B_v,A_{\rm MC},R_{\rm MC}\}$ using the MCMC technique with broad uniform priors is shown in Figs.~\ref{fig:sdbmc-stats-kspace} and~\ref{fig:sdbmc-contours-kspace}. Since we expect the \emph{sdbmc} model to break down at small scales, we restrict the data used in the likelihood calculation to wave numbers smaller than $k_{\rm max}$, whose value we discuss below.

\begin{table}[h]
\centering
\begin{tabular}{cccccc}
\hline\hline
$B_1$ & $B_v$ & $A_{\rm MC}$ & $R_{\rm MC}$  &  $\chi^2/{\rm dof}$ & $p$-value\\
&&& (\Mpch) && \\
\hline 
$2.764$ & $-16.55$ & $0.0230$ & $8.1$  & $41.55/35$ & $0.25$\\
\hline\hline
\end{tabular}
\caption{Best-fit values of the parameters $\{B_1,B_v,A_{\rm MC},R_{\rm MC}\}$ along with the $\chi^2$ per degree of freedom and corresponding $p$-value from the MCMC analysis of the fourier space measurements of \cite{grieb2016} using the \emph{sdbmc} model. See Fig.~\ref{fig:sdbmc-contours-kspace} for the median and central $68\%$ confidence ranges of each parameter in the respective samples. }
\label{tab:sdbmc-bestfit-kspace}
\end{table}

From the \emph{left panel} of Fig.~\ref{fig:sdbmc-stats-kspace} and Table~\ref{tab:sdbmc-bestfit-kspace}, we see that, similarly to the configuration space results in the main text, the \emph{sdbmc} model is capable of producing an excellent description of the simulation measurements at $k<k_{\rm max}$. The \emph{right panel} of Fig.~\ref{fig:sdbmc-stats-kspace} shows the corresponding predictions in configuration space near the BAO feature. Fig.~\ref{fig:sdbmc-contours-kspace} shows that this fit is achieved by a parameter combination in which \emph{all} $4$ parameters are significantly non-zero, with values that are broadly similar to those seen for the $z=0$ configuration space results in the main text. The parameter values can, in principle, depend not only on the tracer selection and epoch of observation but also on the cosmology which is slightly different between the two simulations. The differences between the best fitting \emph{sdbmc} and the \emph{no sdbmc} models are also similar to those discussed in the main text, with the difference between the monopoles being more pronounced.

The results in Figs.~\ref{fig:sdbmc-stats-kspace} and~\ref{fig:sdbmc-contours-kspace} used $k_{\rm max} = 0.13\,\hMpc$. We have checked that increasing $k_{\rm max}$ substantially degrades the goodness-of-fit with steadily changing best fit parameter values, while decreasing it gives a comparable goodness-of-fit and stable best fit parameter values but with wider parameter constraints. This suggests that $k < k_{\rm max}=0.13\,\hMpc$ is a good indicator of the validity of the \emph{sdbmc} model as applied to the measurements from \cite{grieb2016}. It is possible that the value of $k_{\rm max}$ depends on the choice of halo configuration, cosmology, etc., but it suffices for our purposes to note that the \emph{sdbmc} model is an accurate description of anisotropic clustering at \emph{substantially} smaller scales than the model without scale-dependent bias and mode coupling (dashed lines in Fig.~\ref{fig:sdbmc-stats-kspace}).

\section{Accuracy and generality of our treatment of mode coupling}\label{app:mc}

In the main text, we noted that the real space model is `smeared', biased linear theory plus a mode-coupling piece, and the redshift space model has a $\mu$-dependent `Kaiser' pre-factor times $\mu$-dependent smearing.  However, without further comment, this is ambiguous, in that it does not specify if the `bias' factors are `Lagrangian' or `Eulerian', nor how the Kaiser factor arises.  In Lagrangian bias models, there is no ambiguity:  as time goes on, the smearing increases, the bias evolves from Lagrangian to Eulerian, {\rm and} the Kaiser factor is also generated, all self-consistently.  In particular, in the Zeldovich approximation, the redshift-space coordinate $\bm{x}_s$ at late times is related to the initial coordinate $\bm{q}$ by 
\begin{equation}
    \bm{x}_s = \bm{q} + \bm{S}(\bm{q}) + f \bm{S}(\bm{q})\cdot \hat{\bm{z}}
\end{equation}
where $\bm{S}(\bm{q})$ is the gradient of the gravitational potential $\phi(\bm{q})$.
As a result, the redshift-space distorted power spectrum is \cite{dcss10,peaksBAO,iPTpeaks}
\begin{equation}
    \frac{P^s_{\rm Zel}(k,\mu)}{G(k,\mu)} \approx 
    \Big[b_1^{\rm L}(k) + b_{\rm vel}(k)(1 + f\mu^2)\Big]^2\, P_{\rm L}(k) 
    + P^s_{\rm 1loop}(k,\mu),
    \label{eq:PksZel}
\end{equation}
where $\mu$ is the cosine of the angle between vector $\bm{k}$ and the line-of-sight, and 
\begin{align}
    P^s_{\rm 1loop}(k,\mu) &= 
    \int d\bm{k}_1\,P_{\rm L}(k_1)\int d\bm{k}_2\,\,P_{\rm L}(k_2)\,
    \,
    \Big[b_2^{\rm L}(\bm{k}_1,\bm{k}_2) + b_{\rm vel}(\bm{k}_1)b_{\rm vel}(\bm{k}_2) \frac{\bm{K}\cdot\bm{k}_1}{k_1^2}\frac{\bm{K}\cdot\bm{k}_2}{k_2^2}
    \nonumber\\
    &\quad
    + b_1^{\rm L}(k_1)\, b_{\rm vel}(\bm{k}_2)\, \frac{\bm{K}\cdot\bm{k}_2}{k_2^2}
    + b_1^{\rm L}(k_2)\, b_{\rm vel}(\bm{k}_1)\, \frac{\bm{K}\cdot\bm{k}_1}{k_1^2} \Big]^2\,\delta_{\rm D}(\bm{k}-\bm{k}_1-\bm{k}_2) ,
\end{align}
with 
\begin{equation}
    G(k,\mu) \equiv \exp(-[1 + f(2+f)\mu^2]\,k^2\sigv^2)
    \qquad
    {\rm and}\qquad \bm{K}\equiv\bm{k} + f\, (\bm{k}\cdot\hat{\bm{z}})\,\hat{\bm{z}}\,,
\end{equation}
where \sigv\ was defined in \eqn{eq:sigv-def}.
Before proceeding further, notice that this expression shows that the 1-loop contribution is smeared as well.  

For our problem, 
 $b_{\rm vel}(k) = 1 - (s_0/s_1)\,k^2$, 
 $b_1^{\rm L}(k)=b_{10} + b_{01}\,(s_0/s_1)\, k^2$ 
and 
 $b_2^{\rm L}(k_1,k_2) = b_{20} + b_{21}\,(s_0/s_1)\,(k_1^2 + k_2^2) + b_{22}\,(s_0/s_1)^2\,k_1^2 k_2^2$.
This makes 
 $b_1^{\rm L}(k) + b_{\rm vel}(k) = (b_{10}+1) + (b_{01}-1)(s_0/s_1)\,k^2$, 
and provides an easy way to see how the Kaiser factor in the first term on the right hand side of Eq.~(\ref{eq:PksZel}) is modified \cite{peaksRSD}.  When multiplied by $G(k,\mu)$, the first term is the one we called `prop' in the main text.  

The second term is more complicated;  even when there is no $k$-dependent bias, it generates many terms \cite{dcss10}.  However, on the large scales most relevant to BAO analyses, the most relevant terms are those which are proportional to $\mu$ (in the square brackets above) \cite{cs08}.  These are generated by the fact that the initial and final pair separations will be different.  If these differences are small compared to the initial separation, then a Taylor series expansion yields $\delta(\bm{x})\approx \delta(\bm{q}) + (\bm{x}-\bm{q})\cdot\nabla_{\bm{q}}\delta$.  However, $\bm{x}-\bm{q} = \bm{S}$ is just the gradient of the gravitational potential (i.e., the force that produces the acceleration which generates the displacement).  As a result, the second-order correlators which matter are pairwise combinations of $\delta^2$ at one position with $\nabla\phi\cdot\nabla\delta$ at another.  In configuration space, for dark matter, these correspond to the product of a logarithmic derivative of $\xi_{\rm L}$ and a volume integral of $\xi_{\rm L}$.  That product should be convolved with the Gaussian $G(k,\mu)$.  (In principle, this is not the same as convolving each term and then multiplying.)

The volume integral is smoother across the BAO scale than is the derivative; we expect this to be generic.  Therefore, for dark matter, we would follow \citetalias{cs08} in approximating the full mode coupling term as a constant times a term proportional to $d\xi_{\rm L}/d\ln r$.  In addition, to account for the smearing by $G(k,\mu)$, we would only smooth $d\xi_{\rm L}/d\ln r$ (although we should really smooth the product of the two terms), since multiplication by a constant will not affect the smearing.  How does this generalize to biased tracers? 

In the main text, we approximated the full mode coupling term as a constant times a term proportional to $d\xi_{\rm prop}/d\ln r$ (recall that $\xi_{\rm prop}$ denotes the smeared, $k$-dependent, Eulerian-biased, linear theory correlation function).  This has the virtue of not requiring new parameters to specify $b_2(k)$, and would be `exact' for the Lognormal Lagrangian bias model (modulo the fact that we are approximating the other term as a constant, so that the issue of smoothing is trivial).  

\begin{figure}
\centering
\includegraphics[width=0.65\textwidth]{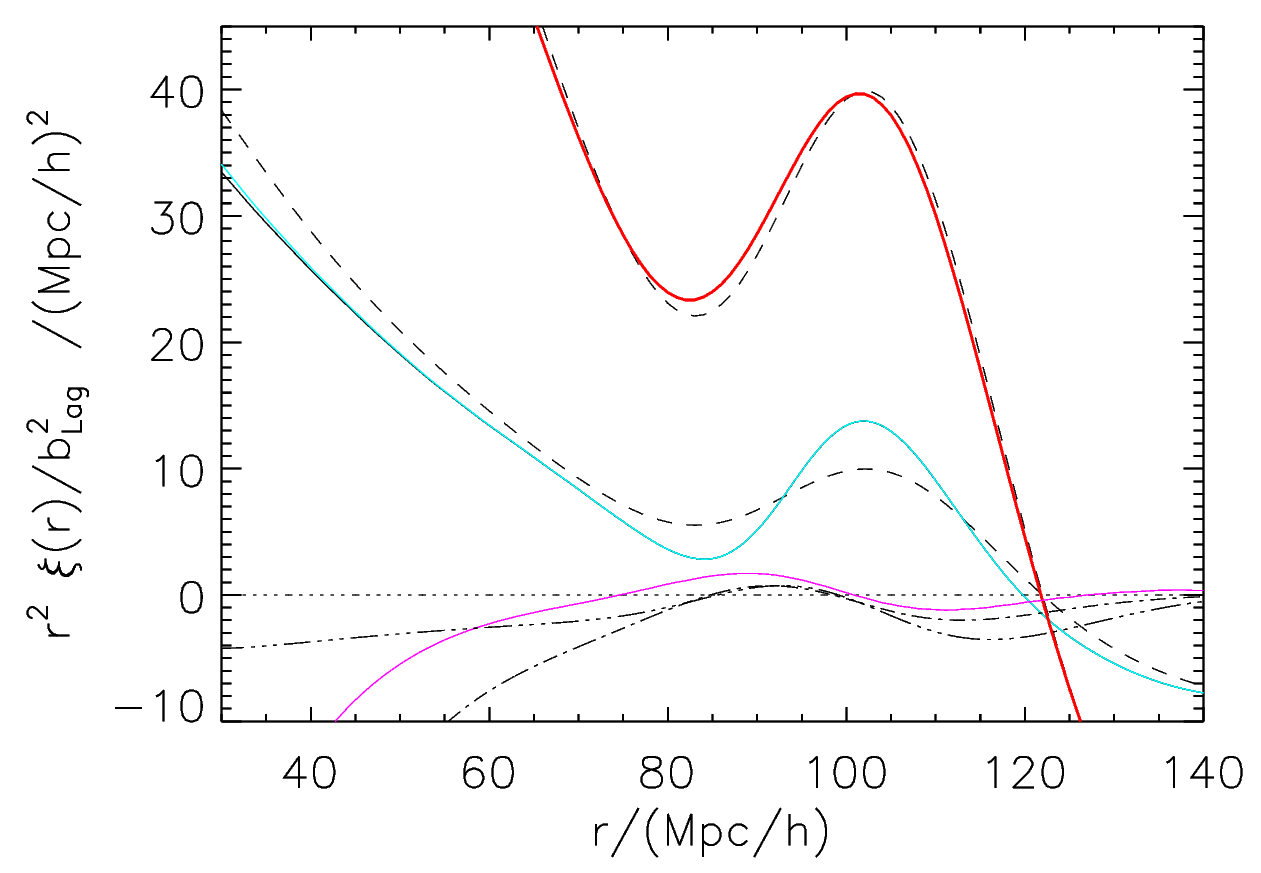}
\caption{Evolution of the shape of the correlation function, and accuracy of our treatment of the mode coupling term.  All curves have been divided by the Lagrangian bias factor $b_{10}^2$.  Solid cyan and red curves show the initial and Zeldovich-evolved correlation functions.  Dashed curves show a smeared version of the initial shape, and this same shape multiplied by $[(b_{10}+1)/b_{10}]^2$ (i.e., $\xi_{\rm prop}/b_{\rm 10}^2$).  Solid magenta curve shows the difference between the red curve and $\xi_{\rm prop}/b_{\rm 10}^2$, and triple-dot dashed curve shows our $\xi_{\rm MC}/b_{\rm 10}^2$ with $\xi_{\rm MC}\propto d\xi_{\rm prop}/d\ln r$.  
}
\label{fig:xiZel-LN}
\end{figure}

Figure~\ref{fig:xiZel-LN} shows that this works reasonably well.  
We assumed the tracers were initially Lognormally biased, with Lagrangian bias $b_{10}=1$.  The cyan solid line shows $r^2\xi_{\rm L}/b_{\rm 10}^2$, and the red solid line shows the associated Zeldovich-evolved correlation function (also normalized by the Lagrangian $b_{10}^2$).  The difference in amplitude is mainly driven by the fact that the bias factor evolves from $b_{10}$ to $b_{10}+1$; there is a secondary effect that arises from the smoothing and mainly alters the significance of the BAO feature.  To illustrate, the lower dashed line shows the result of smearing the linear theory shape, and the upper dashed line shows the result of multiplying this by $[(b_{10}+1)/b_{10}]^2$.  This illustrates that the smeared linear theory, with the Eulerian rather than Lagrangian bias factor -- the quantity called $\xi_{\rm prop}$ in the main text -- is quite a good model for the evolved correlation function.  The magenta curve shows the difference between the solid red and upper dashed curves:  we are interested in how well our approximation for the mode-coupling term accounts for this difference. The triple dot-dashed curve shows our approximation, and the dot-dashed curve accounts for the fact that the volume integral is not quite constant across these scales (an effect we ignore).  Our approximation is quite good -- especially on BAO scales.

We expect our simple parametrization of the `mode coupling' term in the main text to be rather generic, especially because we expect the Zeldovich approximation -- the leading order in any Lagrangian perturbation theory -- to be rather generic.  
Moreover, if we decompose the term in square brackets into a monopole, dipole and quadrupole (for the angle between $\bm{k}_1$ and $\bm{k}_2$, not the angle with respect to the line of sight), then all three terms have coefficient 1/2.  Within GR, when one goes beyond the Zeldovich approximation, then the coefficients of the monopole and quadrupole change, but the dipole does not.  I.e., moving from Zeldovich to the exact dynamics does not change the coefficient of the dipole term.  Likewise, we expect that changing from GR to modified dynamics, at this order, will modify the monopole and quadrupole terms, but will leave the dipole unchanged. However, it is this dipole term which dominates the mode coupling on BAO scales, which is why we believe our treatment of mode coupling is likely to be rather general.  (This dipole term arises from the fact that the displacement is the gradient of the same potential which, when inserted in the Poisson equation, yields the density.  Our mode coupling term will be unrealistic if this is no longer true.  See \cite{mpp24} for a careful discussion of how symmetries, in particular Galilean invariance, constrain LPT and ensure that, to second order, our mode-coupling term is quite general.)  In any case, as the main text shows, the effects of scale-dependent bias are more important than those of mode-coupling, so we believe our simplified treatment is sufficiently accurate.

\end{document}